\documentclass[11pt,aps,nofootinbib,floatfix]{revtex4}
\usepackage{mathrsfs}
\usepackage{graphicx}
\usepackage{amsmath}
\usepackage{amsfonts}
\usepackage{amssymb}
\usepackage{color}

\usepackage{epsfig}
\usepackage{CJK}
\usepackage{graphicx}
\usepackage{epsfig}
\usepackage{eepic}
\usepackage{bbm}
\usepackage{dcolumn}
\usepackage{bm}
\usepackage{ulem}
\usepackage{slashbox}
\usepackage{multirow}

\newcommand{\omits}[1]{}

\def\bc{\begin{center}}
\def\nno{\nonumber}
\def\ec{\end{center}}
\def\be{\begin{eqnarray}}
\def\ee{\end{eqnarray}}

\definecolor{dyellow}{rgb}{1.,0.8,.0}
\definecolor{myblue}{rgb}{.1,.1,.7}
\definecolor{dcyan}{rgb}{.0,.6,.6}
\definecolor{cyan}{rgb}{0.4,1.0,1.0}
\definecolor{dmagenta}{rgb}{0.6,0.0,0.6}
\definecolor{brown}{rgb}{0.6,0.2,0.}
\definecolor{darkblue}{rgb}{.0,.0,0.5}
\definecolor{darkred}{rgb}{0.75,0.0,0.0}
\definecolor{orange}{rgb}{1.,.6,.0}
\definecolor{dorange}{rgb}{0.8,.4,.0}
\definecolor{green}{rgb}{0.0,1.0,0.0}
\definecolor{darkgreen}{rgb}{0.0,0.6,0.0}
\definecolor{purple}{rgb}{.4,.0,.4}
\definecolor{lightgrey}{rgb}{0.7, 0.7, 0.7}
\definecolor{grey}{rgb}{0.4, 0.4, 0.4}

\def\La{\Lambda}
\def\Si{\Sigma}

\def\al{\alpha}
\def\ga{\gamma}
\def\dl{\delta}

\def\la{\lambda}

\def\si{\sigma}
\def\om{\omega}

\def\Na{\nabla}

\newcommand{\nc}{\newcommand}
\nc{\rnc}{\renewcommand} \nc{\ket}[1]{\left | \, #1 \right \rangle}
\nc{\bra}[1]{\left \langle #1 \, \right |}
\nc{\ua}{\uparrow} \nc{\da}{\downarrow}

\nc{\braket}[2]{\langle\, #1\,|\,#2\,\rangle}
\nc{\half}{\frac{1}{2}}

\nc{\prj}{\mathcal{P}} \nc{\hilb}{\mathcal{H}}
\nc{\pth}{\mathcal{C}} \nc{\inprod}[2]{\braket{#1}{#2}}
\nc{\upket}{\ket{\uparrow}} \nc{\downket}{\ket{\downarrow}}
\nc{\upbra}{\bra{\uparrow}} \nc{\downbra}{\bra{\downarrow}}

\begin{document}


\title{Novel Features of the Transport Coefficients in Lifshitz Black Branes}

\author{Jia-Rui Sun$^{1}$} \email{jrsun@ecust.edu.cn}
\author{Shang-Yu Wu$^2$} \email{loganwu@gmail.com}
\author{Hai-Qing Zhang$^3$} \email{hqzhang@cfif.ist.utl.pt}

\affiliation{${}^1$Department of Physics and Institute of Modern
Physics, East China University of Science and Technology, Shanghai
200237, China} \affiliation{${}^2$Institute of Physics,
National Chiao Tung University, Hsinchu 300, Taiwan}
\affiliation{${}^3$CFIF, Instituto Superior {T}\'ecnico,
Universidade {T}\'ecnica de Lisboa, Av. Rovisco Pais 1, 1049-001
Lisboa, Portugal}



\begin{abstract}
We study the transport coefficients, including the  conductivities and shear viscosity of the non-relativistic field
theory dual to the Lifshitz black brane with multiple $U(1)$ gauge
fields by virtue of the gauge/gravity duality. Focusing on the case of double $U(1)$ gauge fields, we systematically investigate the electric, thermal and thermoelectric conductivities for the dual non-relativistic field theory. In the large frequency regime, we find a nontrivial power law behavior in the electric AC conductivity when the dynamical critical exponent $z>1$ in (2+1)-dimensional field theory. The relations between this novel feature and the `symmetric hopping model' in condensed matter physics are discussed. In addition, we also show that the Kovtun-Starinets-Son bound for the shear viscosity to the entropy density is not violated by the additional $U(1)$ gauge fields and dilaton in the Lifshitz black brane.

\end{abstract}


\maketitle

\newpage

\tableofcontents

\section{Introduction}
The holographic principle \cite{'tHooft:1993gx,Susskind:1994vu}, especially with its first realization in string theory-the AdS/CFT correspondence, offers us very intriguing and powerful tools to deal with the strongly coupled quantum systems from the dual viewpoint \cite{Maldacena:1997re,Gubser:1998bc,Witten:1998qj}. The more general framework of the correspondence, which is called the gauge/gravity duality, has been extensively applied to the study of the QCD, Quark Gluon Plasma, hydrodynamics {\it etc.}, for an incomplete list, see \cite{Witten:1998zw,Rey:1998ik,Maldacena:1998im,Erlich:2005qh,Policastro:2001yc,Kovtun:2004de,Brigante:2008gz,Brigante:2007nu,
Policastro:2002se,Kovtun:2003wp,Son:2007vk,Bhattacharyya:2008jc,Bhattacharyya:2008kq,
Hansen:2008tq,Iqbal:2008by,Bredberg:2010ky,Bredberg:2011jq,Compere:2011dx,Cai:2011xv,Brattan:2011my,Huang:2011he,
Niu:2011gu,Eling:2011ct}. In the frame work of the gauge/gravity duality, the features of strongly coupled quantum field theory on the conformally flat boundary can be fully captured by its dual weakly coupled classical gravitational or string theory in the curved bulk spacetime. Even though the gauge/gravity duality is widely believed to be held for arbitrary spacetime backgrounds, so far there are only a few explicit examples, in which the best known one is that the strongly coupled $\mathcal{N}=4$ supersymmetric Yang-Mills theory in four dimensional flat spacetime is equivalent to the classical (weakly coupled) limit of the type IIB superstring theory (supergravity) in AdS$_5$ $\times$ $S^5$ spacetime. For most other cases one still requires the bulk to be asymptotically AdS spacetime whereas the boundary field theory is  conformally invariant and relativistic. However, besides numerous strongly coupled systems in high energy physics described by the relativistic quantum field theory, there also exist large classes of strongly coupled phenomena described by non-relativistic field theory in various condensed matter systems, especially near the (quantum) critical points. Therefore, it is very interesting and important to extend the gauge/gravity duality into a non-relativistic version in order to understand the strongly coupled phenomena in the laboratory condition.

Much progress has been made towards this direction in the past few years. One class of work focused on the study of field theories with the Schr\"{o}dinger symmetry, motivated by the study of fermions at unitarity, see \cite{Son:2008ye,Balasubramanian:2008dm}. Another class of work tried to utilize the dual gravitational theories to study the condensed matter systems near quantum phase transitions that contain the Lifshitz fixed points \cite{Kachru:2008yh,Kovtun:2008qy,Taylor:2008tg,Azeyanagi:2009pr,McGreevy:2009xe,Goldstein:2009cv,
Hartnoll:2009ns,Cheng:2009df,Lemos:2011gy,Ross:2011gu,Fang:2012pw}, such as the strongly correlated electron systems. The particular property of the Lifshitz symmetry is that it consists of the anisotropic scaling
\be x\rightarrow \la x \quad {\rm and}\quad t\rightarrow \la^z t,
\ee
where $z$ is called the {\it dynamical critical exponent}. When $z=1$, the above transformation is the usual relativistic scaling. From the perspective of the gauge/gravity duality, the essential point is to construct bulk gravitational solutions by adding some appropriate sources to realize the boundary non-relativistic quantum field theories with the Lifshitz symmetry. The first attempt was done in \cite{Kachru:2008yh}, in which a four dimensional asymptotic Lifshitz spacetime at zero temperature was obtained in the AdS Einstein gravity together with 1- and 2- form gauge fields. The bulk solution can be viewed as a toy model to provide us some useful descriptions for certain magnetic materials and liquid crystals. Subsequently, many asymptotic Lifshitz black hole solutions have been found and analyzed, see for example \cite{Taylor:2008tg,Goldstein:2009cv,Danielsson:2009gi,Mann:2009yx,Bertoldi:2009vn,Cai:2009ac,Pang:2009pd,Bertoldi:2011zr}. With the help of these solutions, important properties of the dual strongly coupled non-relativisitc quantum field theories, such as the transport coefficients, $n$-point correlation functions, renormalized stress tenosr and higher order corrections \cite{Keranen:2012mx,Pang:2009wa,Ross:2009ar,Mann:2011hg}, can be studied by performing the calculations on the side of the Lifshitz black holes/branes.

The asymptotic Lifshitz solutions can be obtained from different types of theories, the one received much attention is the Einstein-Maxwell-dilaton (EMD) theory, { which can be used to model the dual non-relativistic quantum field theories at finite charge density }. Recently, a class of analytic Lifshitz black hole/brane solutions have been solved in the EMD theory  by adding multiple independent $U(1)$ gauge fields \cite{Tarrio:2011de}. These kinds of charged Lifshitz black hole configurations can provide potential interesting applications to condensed matter systems such as fluids, non-Fermi liquids and conductors that contain the Lifshitz fixed points. Some holographic aspects in these spacetime backgrounds have been brought out, such as the instabilities of dual superfluid by adding probe charged scalar field in the bulk \cite{Mozaffar:2012bp} \footnote{A generalization of theses solutions with additional hyperscaling violation factor was obtained in \cite{Gath:2012pg}, in which their dual nonrelativistic field theories was briefly analyzed as well.}. For other related works, see for example \cite{Tong:2012nf,Alishahiha:2012qu,Gursoy:2012ie}.

The purpose of this paper is to utilize these charged Lifshitz black branes \cite{Tarrio:2011de} to further study certain interesting phenomena of the dual strongly coupled non-relativistic quantum field theory with the Lifshitz fixed points on the boundary. Based on the dictionary of the gauge/gravity duality, we know that the multiple $U(1)$ gauge fields in the bulk will source multiple electric currents in the boundary field theory. As a theoretical model, there is no constraints on the number of independent electric currents even though their physical interpretations are not yet very clear. What we focus in this paper is to investigate the transport coefficients of the dual non-relativistic field theory, which includes the electric conductivity $\sigma$, the thermal conductivity $\bar\kappa$, the thermoelectric conductivity $\alpha$ and the shear viscosity $\eta$. To reach this goal, we consider the linearized gravitational and gauge fields perturbations (the scalar channel and the shear channel) in the bulk EMD theory. In particular,  the bulk Lifshitz black hole can be viewed as the non-relativistic counterpart of the Reissner-Nordstr\"om-AdS black hole when $N=2$. Focusing on this case we calculate the conductivities of the dual non-relativistic field theories numerically, which are expected to capture the universal behavior of a class of conductors near the Lifshitz fixed points. Speicifically, after deriving the renormalized second order on-shell effective action, we work out the numerical results of conductivities, including the electric, thermoelectric and thermal conductivities. In particular, we work in $d=3$ and $d=4$ ($d$ is the dimension of the boundary field theory) for $1\leq z\leq 2$. We find some new frequency dependent power law features of the AC conductivities in the large frequency regime for $1<z\leq2$. The possible relations between these novel features and the `symmetric hopping model' in condensed matter physics are discussed in the context. In addition, an other interesting problem is to see whether these additional bulk $U(1)$ gauge fields and dilaton will affect the famous Kovtun-Starinents-Son (KSS) bound derived in the Einstein gravity \cite{Policastro:2001yc,Kovtun:2004de}. By solving the equation of motion of the transverse graviton at the low frequency limit and applying the linear response theory, we show that this bound is not violated although the additional gauge fields and dilaton do respectivley contribute to the shear viscosity as well as the entropy density of boundary charged fluids.

The outline of the paper is as follows. In Section \ref{sect:intro}, we give a brief review of the Lifshitz black hole/brane backgrounds that we will use in this paper.  In Section \ref{sect:conduc}, we obtain the renormalized second order on-shell action of the perturbations and compute the electric, thermal and thermoelectric conductivities of the boundary non-relativistic field theory, in the $N=2$ case. We calculate the shear viscosity of the boundary fluid both for $N=1$ and generic $N$ cases by solving the equation of motion of the transverse graviton in Section \ref{sect:shear}. Conclusions and discussions are drawn in Section \ref{sect:conclusion}. Besides, we list some detailed calculations for deriving the perturbation equations and the second order on-shell actions in Appendix A.

\section{The configuration of Lifshitz black holes/branes}
\label{sect:intro}
Let us consider the $(d+1)$ dimensional theory {with action} \footnote{{The action in eq.(\ref{EMd1}) is usually referred to as the Einstein-Proca-dilaton (EPD) model when $J(\phi)\neq 0$, i.e. when the gauge field is massive.}}
\begin{equation}\label{EMd1}
I=\int{d^{d+1}x}\sqrt{-g}\left(R-\frac{\gamma(\phi)}{4}F^{2}-\frac{1}{2}(\partial\phi)^{2}-\frac{1}{2}J(\phi)A^{2}-V(\phi)\right),
\end{equation}
where $F_{\mu\nu}=\partial_\mu A_\nu-\partial_\nu A_\mu$ is the $U(1)$ gauge field strength, $\phi$ is the dilaton field, $\ga(\phi)$ is the coupling between the gauge field and the dilaton, $J(\phi)$ is the source term and $V(\phi)$ is the potential term. When we add $N\geq 1$ number of independent $U(1)$ gauge fields, the above variables can be accordingly changed as $\ga(\phi)\rightarrow\sum^{N}_{a=1}\ga_a(\phi)$ and $F_{\mu\nu}\rightarrow F_{a\mu\nu}$, and the equation of motions are
\begin{eqnarray}
\Box\phi &=&\frac{dV(\phi)}{d\phi}+\frac{1}{4}\sum^{N}_{a=1}\frac{d\ga_{a}}{d\phi}F^{2}_a, \nonumber\\
\nabla_{\mu}(f_a(\phi)F_a^{\mu\nu})&=& J A_a^\nu, \nonumber\\
R_{\mu\nu}-\frac{1}{2}g_{\mu\nu}R &=& \frac{1}{2}\sum^{N}_{a=1}\ga_a(\phi)\left(F_{a\mu\la}F_{a\nu}
^{\la}-\frac{1}{4}g_{\mu\nu}F_a^2\right)+\frac{1}{2}J\left(A_{a\mu}A_{a\nu}-\frac 1 2 A_a^2 g_{\mu\nu}\right)\nno\\&&+\frac{1}{2}\left(\partial_{\mu}\phi\partial_{\nu}\phi-\frac{1}{2}g_{\mu\nu}(\partial\phi)^{2}-g_{\mu\nu}V(\phi)\right).
\end{eqnarray}
The Lifshitz black holes can be obtained from the following ansatz
\be\label{metricansatz}
ds^2 &=& -\xi(r) e^{-\chi(r)}dt^{2}+\frac{dr^{2}}{\xi(r)}+b^{2}(r)dx^i dx_i,\\\nno
 A_a&=&A_{at}(r)dt\ee
together with
\be
 \quad \phi=\phi(r),\quad J(\phi)=0,\quad V(\phi)=2\La \quad {\rm and}\quad \ga_a = e^{\lambda_{a}\phi}.\ee
Note that now the EPD model becomes the EMD model since we have set $J(\phi)=0$. 

For $N=1$ case, the solution is
\begin{eqnarray}\label{N=1}
&&ds^{2}=-\frac{r^{2z}}{l^{2z}}\Xi(r)dt^{2}+\frac{l^2 dr^{2}}{r^{2}\Xi(r)}+\frac{r^{2}}{l^2}\sum^{d-1}_{i=1}dx_{i}^{2}, \nonumber\\
&&\Xi(r)=1-\frac{r^{z+d-1}_h}{r^{z+d-1}}, \quad A'_{1t}=l^{-z}\sqrt{2(d+z-1)(z-1)}\mu^{\sqrt{\frac{d-1}{2(z-1)}}}r^{d+z-2}, \quad e^{\phi}=\mu r^{\sqrt{2(z-1)(d-1)}},\nonumber \\
&&\lambda=-\sqrt{\frac{2(d-1)}{z-1}}, \quad
\Lambda=-\frac{(z+d-2)(z+d-1)}{2l^{2}},
\end{eqnarray}
where $l$ is the curvature radius of the Lifshitz spacetime, $\mu$ is the scalar field amplitude, $m$ is related to the mass of the black hole and ``$'$" is the derivative with respect to $r$. The Hawking temperature and the Bekenstein-Hawking entropy are respectively
\begin{equation}
T=\frac{(z+d-1)r^{z}_h}{4\pi l^{z+1}}, \quad
S_{BH}=\frac{V_{d-1}}{4G_{d+1}}\left(\frac{r_h}{l}\right)^{d-1},
\end{equation}
and $V_{d-1}=\int d^{d-1}x$ is the spatial volume of the boundary.

For generic $N$, the black hole solution is \cite{Tarrio:2011de}
\begin{eqnarray}\label{bgd}
&&ds^{2}=-\frac{r^{2z}}{l^{2z}}f_k(r)dt^{2}+\frac{l^{2}}{r^{2}f_k(r)}dr^{2}+\frac{r^{2}}{l^2}d\Omega^{2}_{k,d-1}, \nonumber\\
&&f_k(r)=k(\frac{d-2}{d+z-3})^{2}\frac{l^{2}}{r^{2}}+1-mr^{-(d+z-1)}+\sum^{N-1}_{a=2}\frac{\rho^{2}_a\mu^{-\sqrt{2\frac{z-1}{d-1}}}l^{2z}}{2(d-1)(d+z-3)}r^{-2(d+z-2)}, \nonumber\\
&&A'_{1t}=l^{-z}\sqrt{2(d+z-1)(z-1)}\mu^{\sqrt{\frac{d-1}{2(z-1)}}}r^{d+z-2},\nonumber\\
&&A'_{at}=\frac{\rho_{a}\mu^{-\sqrt{2\frac{z-1}{d-1}}}}{r^{d+z-2}}, \quad (a=2,\ldots,N-1)\nonumber\\
&&A'_{Nt}=l^{1-z}\frac{\sqrt{2k(d-1)(d-2)(z-1)}}{\sqrt{d+z-3}}\mu^{\frac{d-2}{\sqrt{2(d-1)(z-{ 1})}}}r^{d+z-4},\nonumber\\
&&\lambda_1=-\sqrt{\frac{2(d-1)}{z-1}},\quad \lambda_a=-\sqrt{\frac{2(z-1)}{d-1}},\quad \la_N=-\frac{d-2}{d-1}\sqrt{\frac{2(d-1)}{z-1}},\quad (a=2,\ldots,N-1), \nno\\
&&e^{\phi}=\mu r^{\sqrt{2(d-1)(z-1)}},\quad \Lambda=-\frac{(d+z-1)(d+z-2)}{2l^{2}},
\end{eqnarray}
where $\rho_a$ are related to the charges of the black hole, while $k$ is the factor indicating the topology of the horizon.  For $k=0$, the horizon is flat; for $k=-1$, the horizon is hyperbolic and the horizon is spherical for $k=1$. In the following, we shall take the spatial flat case, namely, the Lifshitz black brane with $k=0$. When $N\geq 2$, the black brane will contain multiple horizons in the presence of electromagnetic fields, let us define the outer event horizon to be located at $r=r_h$, i.e.
$f(r_h)=0$. Then the temperature of the black brane is
\be
T=\frac{1}{4\pi}\left(\frac{r_h}{l}\right)^{z+1}f'(r_h)=\frac{1}{4\pi}\left(\frac{r_h}{l}\right)^{z+1}\left(\frac{2(d+z-2)}{r_h}-\frac{m(d+z-3)}{r_h^{d+z}}\right),
\ee
where
$f(r)=1-mr^{-(d+z-1)}+\sum^{N-1}_{j=2}\frac{\rho^{2}_{j}\mu^{-\sqrt{2\frac{z-1}{d-1}}}l^{2z}}{2(d-1)(d+z-3)}r^{-2(d+z-2)}$. The horizon entropy $S_{BH}$ and entropy density $s$ of the dual
CFT are
\be S_{BH}=\frac{r_h^{d-1}}{4G_{d+1}l^{d-1}}V_{d-1}\quad {\rm
and}\quad
s=\frac{S_{BH}}{V_{d-1}}=\frac{r_h^{d-1}}{4G_{d+1}l^{d-1}},\ee

\section{The conductivities}
\label{sect:conduc}
In this section, we will compute the conductivities of the non-relativistic quantum field theory dual to the Lifshitz black brane. The electric conductivity $\sigma$ can be calculated by just turning on the bulk gauge field fluctuations $\delta A_{x}(t,r)=a_{x}(r)e^{-i\omega t}$. However, if we want to consider the thermal conductivity $\bar\kappa$ and the thermoelectric conductivity $\al$ , we need to consider the back reaction of the gauge fields to the metric, namely, we need to meanwhile turn on $\delta
g_{tx}(t,r)=h_{tx}(r)e^{-i\omega t}$. For the EMD theory (when taking $J=0$) in eq.(\ref{EMd1}),
we can obtain the linearized Einstein and Maxwell equations as (see Appendix A for details)
\be\label{lEM2} h'_{tx}-\frac{2b'}{b}h_{tx}+\sum_{a=1}^N
\gamma_a(\phi)A'_{a t}a_{a x}&=&0,\\
\label{lM2}
a''_{ax}+\left(\frac{(d-3)b'}{b}+\frac{\xi'}{\xi}-\frac{\chi'}{2}+\frac{\phi'}{\gamma_a(\phi)}
\frac{d\gamma_a(\phi)}{d\phi}\right)a'_{ax}
+\frac{\omega^2}{\xi^2}e^{\chi}a_{ax}&=&\left(\frac{2b'h_{tx}}{\xi
b}-\frac{h'_{tx}}{\xi}\right)A'_{at}e^{\chi}.\ee
where $b(r), \xi(r)$ and $\chi(r)$ are factors in eq.\eqref{metricansatz}.
Note that eq.(\ref{lEM2}) is the first order differential equation for $h_{tx}$ which can be integrated out as
\be h_{tx}=-b(r)^2\int \frac{1}{b(r)^2}\sum_{a=1}^N
\gamma_a(\phi)A'_{a t}a_{a x}dr,\ee
and eq.(\ref{lM2}) can be written into the following equation as
\be\label{ax}
a''_{ax}+\left(\frac{(d-3)b'}{b}+\frac{\xi'}{\xi}-\frac{\chi'}{2}+\frac{\phi'}{\gamma_a(\phi)}
\frac{d\gamma_a(\phi)}{d\phi}\right)a'_{ax}
+\frac{\omega^2}{\xi^2}e^{\chi}a_{ax}=\frac{1}{\xi}\left(\sum_{c=1}^N
\gamma_c(\phi)a_{c x}A'_{c t}\right)A'_{at}e^{\chi}\ee
with the help of eq.(\ref{lEM2}).

When $N=1$, the background Lifshitz black brane eq.\eqref{N=1} is neutral as the Schwarzschild AdS black brane, the electric conductivity has been studied by adding a probe $U(1)$ gauge field in the bulk in \cite{Pang:2009wa}.

In the following, we will focus on the $N=2$ situation, in which
\be\label{N2sol}
&&e^{-\chi}=\left(\frac{r}{l}\right)^{2z-2}, \quad
\xi(r)=\frac{r^{2}}{l^{2}}f(r),
\quad e^{\phi}=\mu r^{\sqrt{2(d-1)(z-1)}},\quad b(r)=\frac{r}{l},\nno\\
&&f(r)=1-mr^{-(d+z-1)}+\frac{\rho^{2}_{2}\mu^{-\sqrt{2\frac{z-1}{d-1}}}l^{2z}}{2(d-1)(d+z-3)}r^{-2(d+z-2)}.\ee
Recall that for the $N=2$ case, the background gauge field $A_{1t}$ is divergent at the spatial infinity, it only supports the asymptotic Lifshitz geometry instead of contributing to the free charge of the background electromagnetic field \cite{Tarrio:2011de}. On the contrary, the gauge field $A_{2t}$ plays the role of the free electromagnetic field. Besides, our numeric results show that the asymptotic expansion of $a_{1x}$ is also divergent at the spatial infinity. Thus only the fluctuations of $A_{2}$, namely, $a_{2x}$ is the genuine electromagnetic perturbations, which will contribute to the electric conductivities of the dual field theory on the boundary. Consequently, to study the conductivities, we only need to turn on the perturbations $a_{2x}$ and $h_{tx}$, while turning off the perturbation $a_{1x}$. Then after substituting the above black brane solution eq.(\ref{N2sol}) into the original fluctuation equations (\ref{lEM2}) and (\ref{ax}), we obtain
 \be
 h_{tx}'-\frac2r h_{tx}+\rho_2r^{z-d}a_{2x}&=&0,\\
\label{eoma2x} a''_{2x}+\left(\frac{f'}{f}+\frac{d+3z-4}{r}\right)a'_{2x}+\left(\frac{\om^2l^{2z+2}}{f^2r^{2z+2}}-\frac{\rho_2^2\mu^{-\sqrt{2\frac{z-1}{d-1}}}r^{(2-2d-2z)}l^{2z}}{f}\right)a_{2x}&=&0.
  \ee
The explicit asymptotic behavior of $a_{2x}$ near the infinite boundary with certain $d$ and $z$ considered in this paper can be found in Table \ref{expansion}, in which $C_1$ and $C_2$ are expansion coefficients that depend on the frequency $\omega$. According to the gauge/gravity duality, $C_1$ represents the source while $C_2$ represents the vacuum expectation value of the current operator $\mathfrak{J}_x$ dual to $a_{2x}$.
 \begin{table}[h]
 {\large
 \begin{tabular}{|c||c|c|c|}
       \hline
          &$z=1$ & $z=3/2$ & $z=2$  \\
        \hline\hline
       $d=3$ & $C_1+\frac{C_2}{r}$ & $C_1+\frac{C_2}{r^{5/2}} $&$ C_1+\frac{C_1\om^2\log(r)}{4r^4}+\frac{C_2}{r^4}$  \\
       \hline
       $d=4$ & $C_1+\frac{C_1\om^2\log(r)}{2r^2}+\frac{C_2}{r^2}$ & $C_1+\frac{2C_1\om^2}{3r^3}+\frac{C_2}{r^{7/2}}$ & $C_1+\frac{C_1\om^2}{4r^4}+\frac{C_2}{r^5}$  \\
       \hline
       \end{tabular}
       \caption{\label{expansion} The expansions of $a_{2x}$ with respect to various $d$ and $z$ near infinity. The coefficients $C_1$ and $C_2$ are functions of the frequency $\om$.}
       }
       \end{table}

 In addition, the asymptotic behavior of $h_{tx}$ near the infinity boundary is,
 \be h_{tx}\sim r^2 h_{tx}^{(0)}+\frac{h_{tx}^{(1)}}{r^{(d-z-1)}}+\cdots,\ee
where, $h_{tx}^{(1)}=C_1\rho_2/(1+d-z)$ in which $C_1$ is the source term of the expansions in $a_{2x}$, see Table \ref{expansion}.

\subsection{Second order on-shell action}
 In order to compute the transport coefficients of $\sigma, \al$ and $\bar{\kappa}$, we need to know the quadratic on-shell actions for these perturbations. The on-shell action for the perturbation $a_{2x}$ and $h_{tx}$ up to 2nd order is (we have set $l=1$),
 \be\label{onshell}
 S^{(2)}_{\text{on-shell}}=S^{(2)}_{a_{2x}}+S^{(2)}_{h_{tx}}, \ee
 where
 \be
 \label{onshella2x}S^{(2)}_{a_{2x}}&=&\int d^dx\bigg(-\frac12a_{2x}a'_{2x}e^{-\chi/2}\ga_2(\phi) \xi r^{d-3}\bigg)\bigg|_{r\rightarrow\infty},\\
 \label{onshellhtx}S^{(2)}_{h_{tx}}&=&\int d^dxe^{\chi/2}r^{d-3}\left(-h_{tx}h_{tx}'+\frac12h_{tx}^2(\frac{\xi'}{\xi}-\chi')\right)\bigg|_{r\rightarrow\infty}.
 \ee
  Usually the on-shell action eq.\eqref{onshell} is divergent near the asymptotic boundary, the divergence can be eliminated through the holographic renormalization approach, i.e. by adding appropriate boundary counter terms to the action (see, for example \cite{Balasubramanian:1999re,Skenderis:2002wp, Tarrio:2012xx}). In the configuration of the Lifshitz black brane, the counter terms have different forms with respect to different $d$ and $z$. We will list them in the following:

First of all, we will introduce the counter terms to $S^{(2)}_{a_{2x}}$ in eq.\eqref{onshella2x}. These counter terms are classified according to the expansions of $a_{2x}$ in Table \ref{expansion}.

(a). $\mathbf{d=3,z=1~and ~3/2}$:

In this case, the on-shell action of $S^{(2)}_{a_{2x}}$ is finite at the infinite boundary. There are no counter terms to $S^{(2)}_{a_{2x}}$ just like in the usual relativistic holographic superconductors \cite{Hartnoll:2008kx}.

(b). $\mathbf{(d=3,z=2) ~and~ (d=4,z=1)}:$

For $(d=3, z=2)$ and $(d=4, z=1)$, there will be logarithmic divergence for $S^{(2)}_{a_{2x}}$ on the infinite boundary. In this case, the generic expansions of $a_{2x}$ near $r\rightarrow\infty$ now is,
\be
a_{2x}(r)\sim C_1+\frac{C_1\om^2\log(r)}{(d+3z-5)r^{d+3z-5}}+C_2\left(\frac1r\right)^{d+3z-5}.
\ee
The divergent term of $S^{(2)}_{a_{2x}}$ can be obtained from eq.\eqref{onshella2x} as,
\be
I_{\text{div}.a_{2x}}=\frac{V_{d-1}}{T}{C_1}^2 \om^2  \log (r) \mu^{\sqrt{2\frac{z-1}{d-1}}}
\ee
where $T$ is the temperature of the boundary field theory and $\frac{V_{d-1}}{T}$ is just the volume integration $\int dt d^{d-1}x_i$. Therefore, in this case the counter term should be,
\be
I_{\text{ct}.a_{2x}}=-\frac{1}{2}\log(r)\int d^dx\sqrt{-\ga^0}~\ga_2(\phi)(F^0_{ij})^2.
\ee
where, $\ga^0$ is the determinant of the induced metric while $F^0_{ij}$ is the induced gauge field strength on the asymptotic UV cut off boundary, respectively. It is easy to get that $(F^0_{ij})^2=2\om^2 (a_{2x})^2 r^{-2z}/\xi$. Therefore, the finite on-shell $S^{(2)}_{a_{2x}}$ is,
\be
I_{a_{2x}}^{(2)}=S^{(2)}_{a_{2x}}+I_{\text{ct}.a_{2x}}=\int d^dx ~\left(C_1C_2(d+3z-5)-\frac{C_1^2\om^2}{d+3z-5}\right).
   \ee

(c). $\mathbf{d=4,z=3/2~ and ~ 2}$:

For $d=4$, $z=3/2$ and $z=2$, the general expansions of $a_{2x}$ is,
\be
a_{2x}(r)\sim C_1+\frac{C_1 \om^2 }{2 (z-1) z}\left(\frac{1}{r}\right)^{2 z}+C_2   \left(\frac{1}{r}\right)^{3 z-1}.
   \ee
In this case, the divergent term of $S^{(2)}_{a_{2x}}$ is ,
\be
I_{\text{div}.a_{2x}}=\frac{V_{d-1}}{T}\frac{\om^2\mu^{\sqrt{2(z-1)/3}}C_1^2}{z-1}r^{z-1},
\ee
The counter term for this divergence now is,
\be I_{ct.a_{2x}}=\frac{-1}{2z-2}\int d^dx\sqrt{-\ga^0}~\ga_2(\phi)(F^0_{ij})^2. \ee
Therefore, from the expansions, we can get the finite on-shell action as,
\be
I_{a_{2x}}^{(2)}=S^{(2)}_{a_{2x}}+I_{ct.a_{2x}}=\int d^dx ~C_1 C_2 (3 z-1) \mu ^{\sqrt{2(z-1)/3}}.
\ee

Next, we will introduce the counter terms for the on-shell action $S^{(2)}_{h_{tx}}$ in eq.\eqref{onshellhtx}. We can expand it near $r\rightarrow\infty$ as,
\be S^{(2)}_{h_{tx}}=I_{\text{div}.h_{tx}}+I_{\text{finite}.h_{tx}},\ee
where,
\be
\label{Idiv}I_{\text{div}.h_{tx}}&=&\int d^dx ~(h_{tx}^{(0)})^2 (z-2) r^{d-z+1},\\
\label{Ifin}I_{\text{finite}.h_{tx}}&=&\int d^dx ~\left(\frac{
   (d+z-3)}{d-z+1}C_1 \rho _2h_{tx}^{(0)}+\frac{1}{2} (h_{tx}^{(0)})^2 m (d+z-1) r^{2-2 z}\right).\ee
It can be found that when $z=1$ the last term in $I_{\text{finite}.h_{tx}}$ is finite while for $z>1$ it will vanish at $r\rightarrow \infty$.
As usual, we can introduce the Gibbons-Hawking term $I_{GH}$ and a counter term for the cosmological constant $I_{ct.cc}$ into the on-shell action to cancel the divergence, \footnote{ {The counter terms for the Lifshitz spacetime in eq.(10) in the paper \cite{Tarrio:2012xx} will be the same as ours if they restricted to the Ricci flat boundary.} } where
\be
\label{IGH}I_{GH}&=&2\int d^dx \sqrt{-\ga^0}K, \\
\label{Icc}I_{ct.cc}&=&2\int d^dx \sqrt{-\ga^0}(d-1).\ee
in which, $K=\ga^0_{\mu\nu}\nabla^\mu n^\nu$ is the trace of the extrinsic curvature while $n^\mu$ is the outward pointing unit normal vector on the boundary. Expanding eq.\eqref{IGH} and eq.\eqref{Icc} to the quadratic order of the perturbations near $r\rightarrow \infty$, we arrive at,
\be
\label{IGH2}I_{GH}^{(2)}&=&\int d^dx ~\left((h_{tx}^{(0)})^2 (z-d-1) r^{d-z+1}+\frac{1}{2} (h_{tx}^{(0)})^2 m (d+z-1) r^{2-2 z}\right),\\
\label{Icc2}I_{ct.cc}^{(2)}&=&\int d^dx ~\left((h_{tx}^{(0)})^2 (d-1) r^{d-z+1}+\frac{2  (d-1) }{d-z+1}C_1\rho _2 h_{tx}^{(0)}+\frac{1}{2} (h_{tx}^{(0)})^2 (d-1) m r^{2-2 z}\right).
\ee
Therefore, the total finite on-shell action of the perturbation $h_{tx}$ can be obtained from eqs.\eqref{Idiv}, \eqref{Ifin}, \eqref{IGH2} and \eqref{Icc2} as,
\be
I_{h_{tx}}^{(2)}&=& S^{(2)}_{h_{tx}}-I_{GH}^{(2)}-I_{ct.cc}^{(2)}\\
&=&\int d^dx ~\left(-C_1\rho_2h_{tx}^{(0)}-\frac{m(d-1)}{2}(h_{tx}^{(0)})^2r^{2-2z}\right)\\
&=&\int d^dx ~\left( -(d+1-z)h_{tx}^{(0)}h_{tx}^{(1)}-\frac{m(d-1)}{2}(h_{tx}^{(0)})^2r^{2-2z}\right).
\ee

So, finally the total renormalized quadratic on-shell action for the perturbations $a_{2x}$ and $h_{tx}$ is,
\be
I^{(2)}_{\text{total}}=I_{h_{tx}}^{(2)}
=\int d^dx ~\left(C_1C_2(d+3z-5)-(d+1-z)h_{tx}^{(0)}h_{tx}^{(1)}-\frac{m(d-1)}{2}(h_{tx}^{(0)})^2r^{2-2z}\right),
\ee
for $d=3, z=1~\text{and}~3/2$;
 Or,
\be
I^{(2)}_{\text{total}}&=&I_{a_{2x}}^{(2)}+I_{h_{tx}}^{(2)}\nno\\
&=&\int d^dx ~\left(C_1C_2(d+3z-5)-\frac{C_1^2\om^2}{d+3z-5} -(d+1-z)h_{tx}^{(0)}h_{tx}^{(1)}-\frac{m(d-1)}{2}(h_{tx}^{(0)})^2r^{2-2z}\right),~~~~~
\ee
when $(d=3, z=2)$ and $(d=4, z=1)$; Or,
\be
I^{(2)}_{\text{total}}&=&I_{a_{2x}}^{(2)}+I_{h_{tx}}^{(2)}\nno\\
&=&\int d^dx ~\left(C_1 C_2 (3 z-1) \mu ^{\sqrt{2(z-1)/3}} -(d+1-z)h_{tx}^{(0)}h_{tx}^{(1)}-\frac{m(d-1)}{2}(h_{tx}^{(0)})^2r^{2-2z}\right).
\ee
when $d=4, z=3/2 ~\text{and} ~2$.

\subsection{The electric, thermoelectric and thermal conductivities}
As long as we get the quadratic on-shell action for the perturbations, we can derive the electric and thermal transport coefficients jointly as follows:
\be
      \left(
             \begin{array}{c}
             { \langle} \mathfrak{J}_x { \rangle}\\
              { \langle}  Q_x { \rangle}\\
             \end{array}
           \right)=\left(
                     \begin{array}{cc}
                       \sigma & \al T \\
                       \al T & \bar{\kappa}T \\
                     \end{array}
                   \right) \left(
                             \begin{array}{c}
                               E_x \\
                               -(\nabla_xT)/T \\
                             \end{array}
                           \right),
\ee
where $\mathfrak{J}_x$ is the electric current and $Q_x$ is the heat current, both are in the $x$-direction. And $\sigma, \al$ and $\bar\kappa$ are the electric conductivity, the thermoelectric conductivity and the thermal conductivity, respectively. Following the procedures in \cite{Hartnoll:2008kx,Hartnoll:2009sz}, we can obtain these transport coefficients which are listed in Table \ref{tracoe}:

\begin{table}[h]
{\large
 \begin{tabular}{| c | c || c | c | c |}
       \hline
       \multicolumn{2}{|c||}{}  &$\sigma$ & $\al$ & $\bar\kappa$  \\
        \hline\hline
        \multirow{3}{*}{$d=3$} &$z=1$ & $\frac{C_2(d+3z-5)}{i\om C_1}$ & $-\frac{\rho_2}{i\om T}-\frac{\mu\sigma}{T} $&$ -\frac{m(d-1)}{i\om T}+\mu^2\sigma$  \\
        &$z=\frac32$ & $\frac{C_2(d+3z-5)}{i\om C_1}$ & $-\frac{\rho_2}{i\om T}-\frac{\mu\sigma}{T} $&$\mu^2\sigma$  \\
        &$z=2$ & $\frac{C_2(d+3z-5)}{i\om C_1}-\frac{2\om}{i(d+3z-5)}$ & $-\frac{\rho_2}{i\om T}-\frac{\mu\sigma}{T}$&$ \mu^2\sigma$  \\
       \hline\hline
       \multirow{3}{*}{$d=4$} &$z=1$ & $\frac{C_2(d+3z-5)}{i\om C_1}-\frac{2\om}{i(d+3z-5)}$ & $-\frac{\rho_2}{i\om T}-\frac{\mu\sigma}{T} $&$ -\frac{m(d-1)}{i\om T}+\mu^2\sigma$  \\
        &$z=\frac32$ & $\frac{C_2(3z-1)\mu^{\sqrt{2(z-1)/3}}}{i\om C_1}$ & $-\frac{\rho_2}{i\om T}-\frac{\mu\sigma}{T} $&$ \mu^2\sigma$  \\
        &$z=2$ & $\frac{C_2(3z-1)\mu^{\sqrt{2(z-1)/3}}}{i\om C_1}$ & $-\frac{\rho_2}{i\om T}-\frac{\mu\sigma}{T} $&$\mu^2\sigma$  \\
       \hline
       \end{tabular}
       \caption{\label{tracoe} The various conductivities for different $z$ and $d$.}
 }
       \end{table}

\begin{figure}[h]
  \includegraphics[scale=0.6]{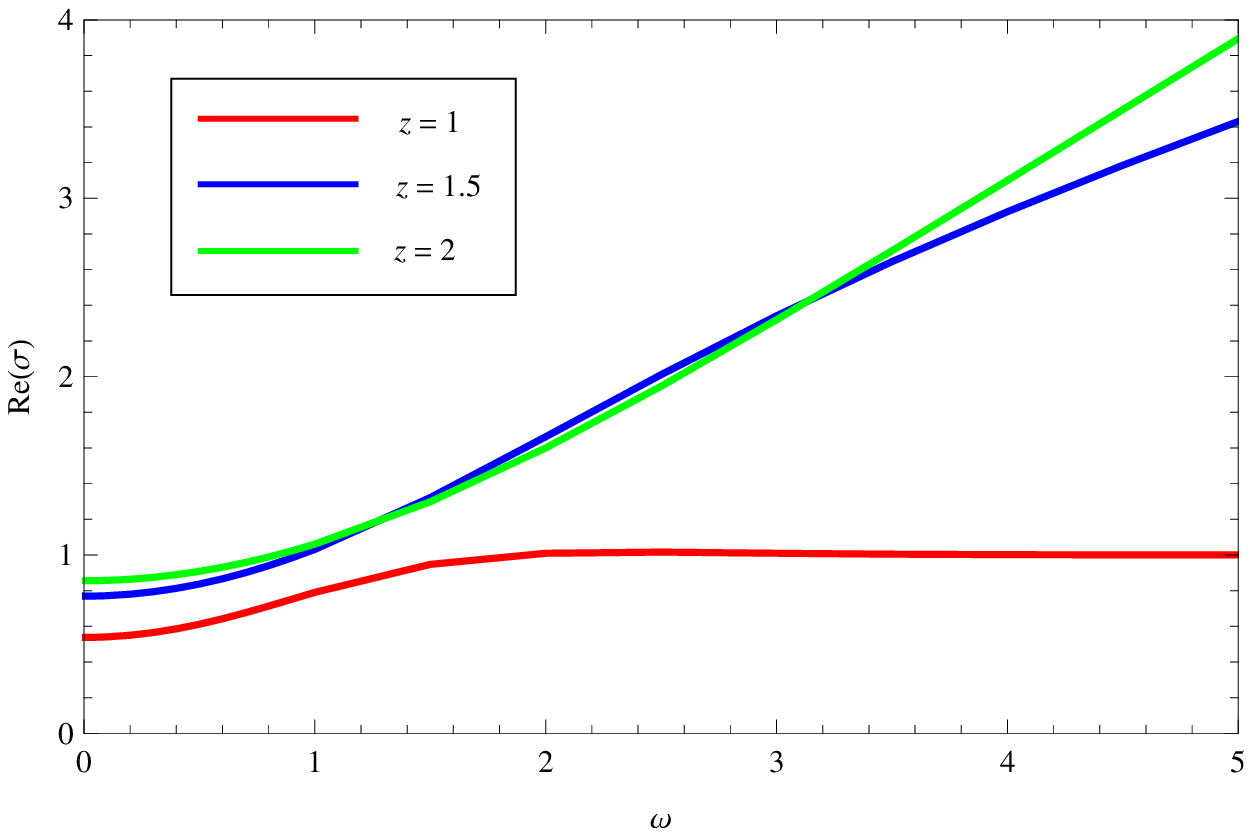}
  \includegraphics[scale=0.6]{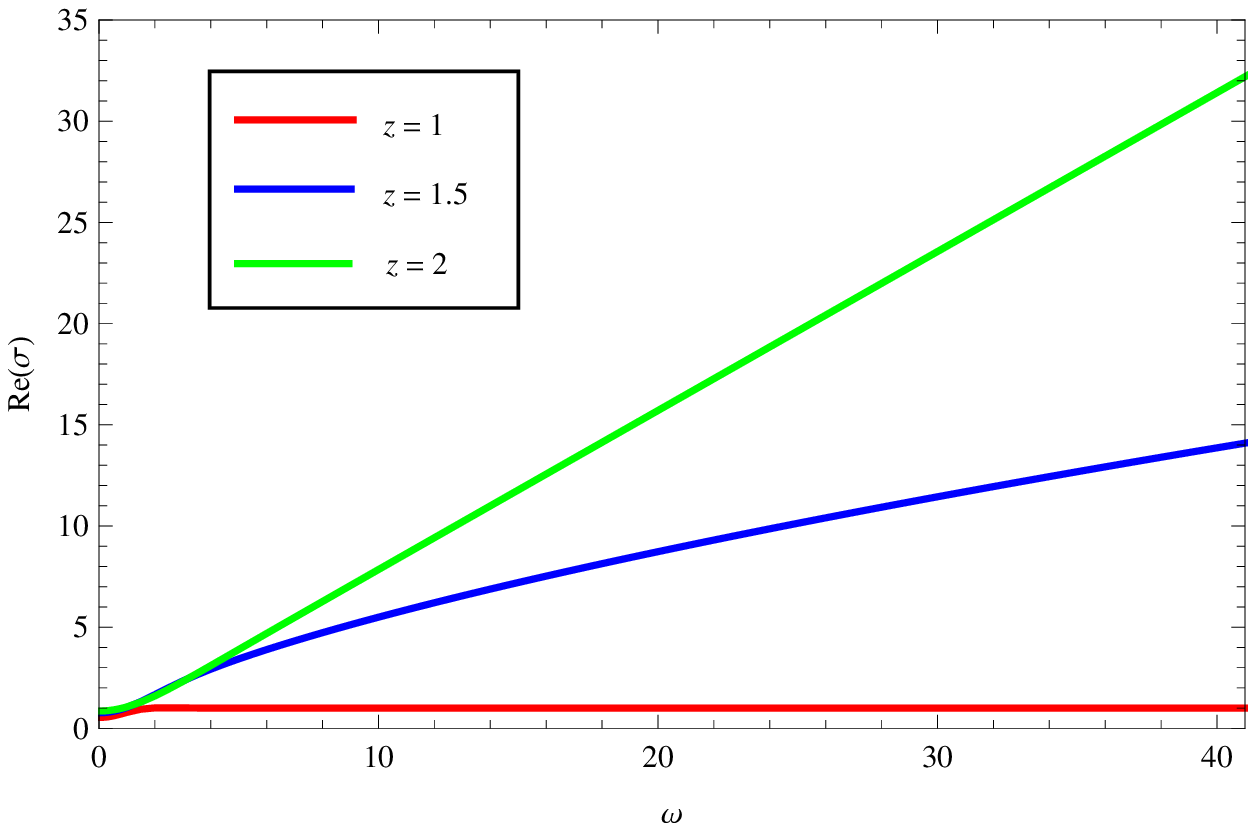}
   \includegraphics[scale=0.6]{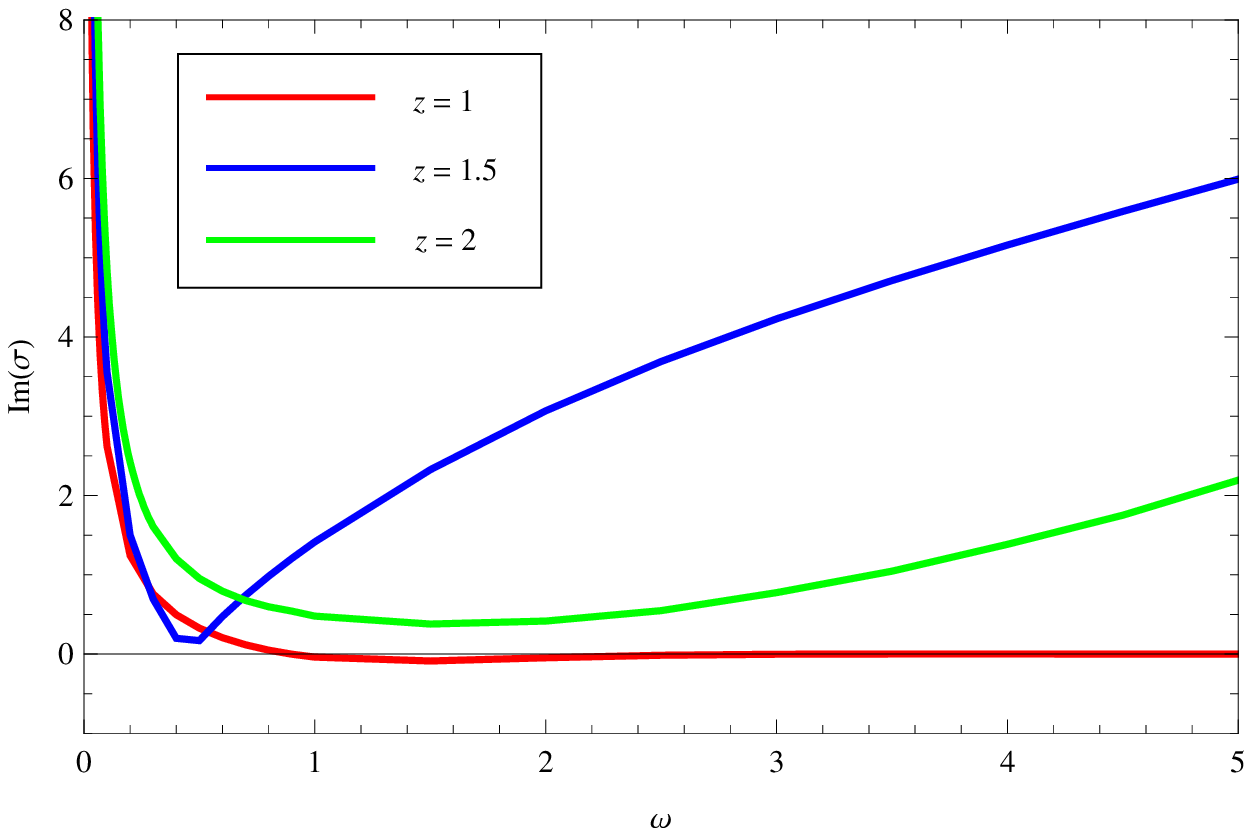}
  \includegraphics[scale=0.6]{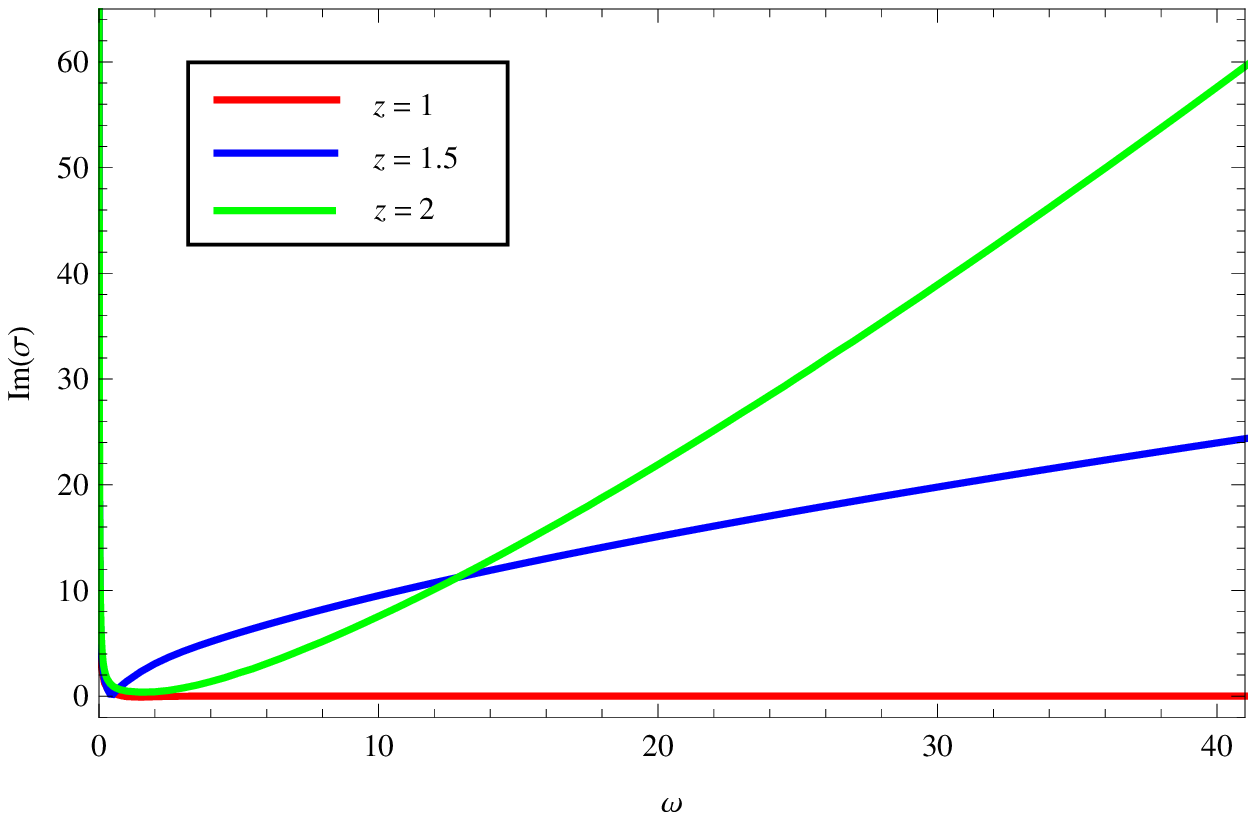}
  \caption{\label{d3}The real and imaginary parts of the conductivity for $d=3$ with respect to various $z$. The left parts are of the low frequency regime while the right parts are of the high frequency regime.}
\end{figure}

\begin{figure}[h]
  \includegraphics[scale=0.6]{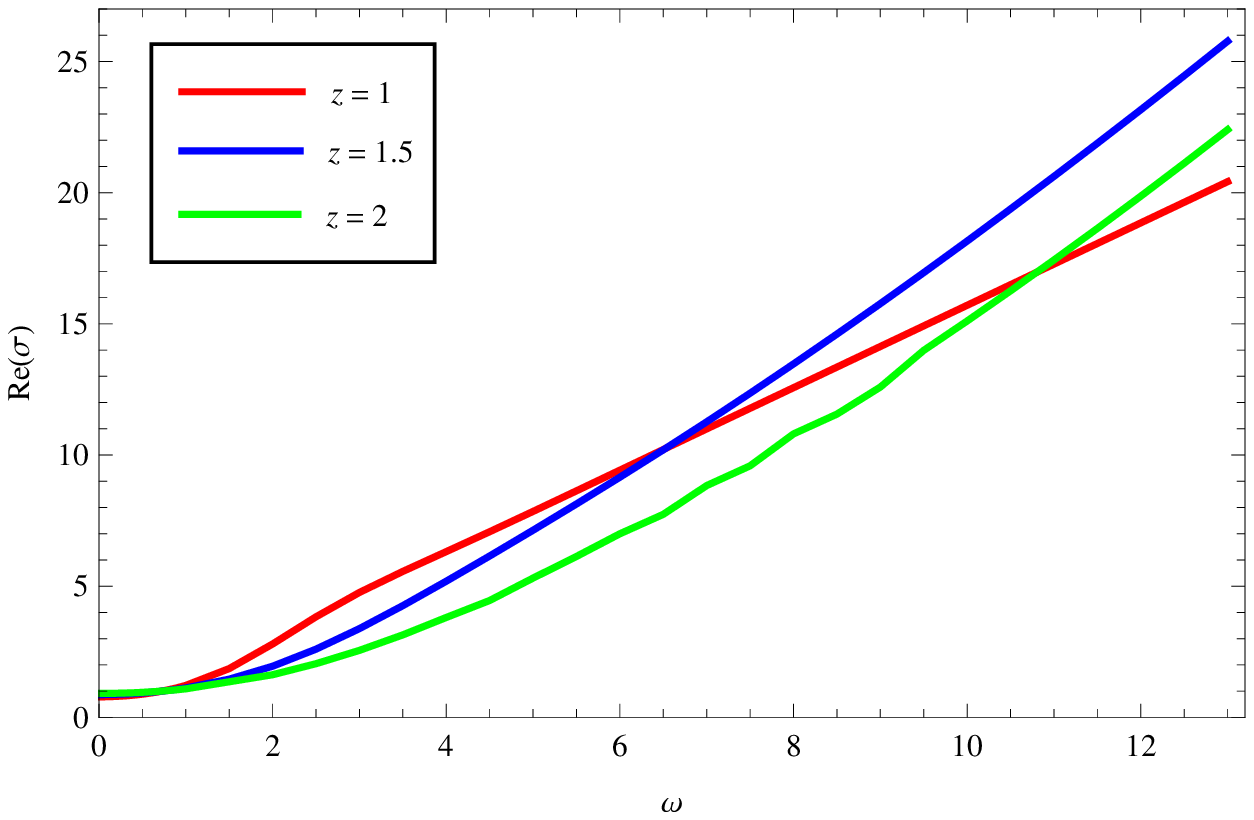}
  \includegraphics[scale=0.6]{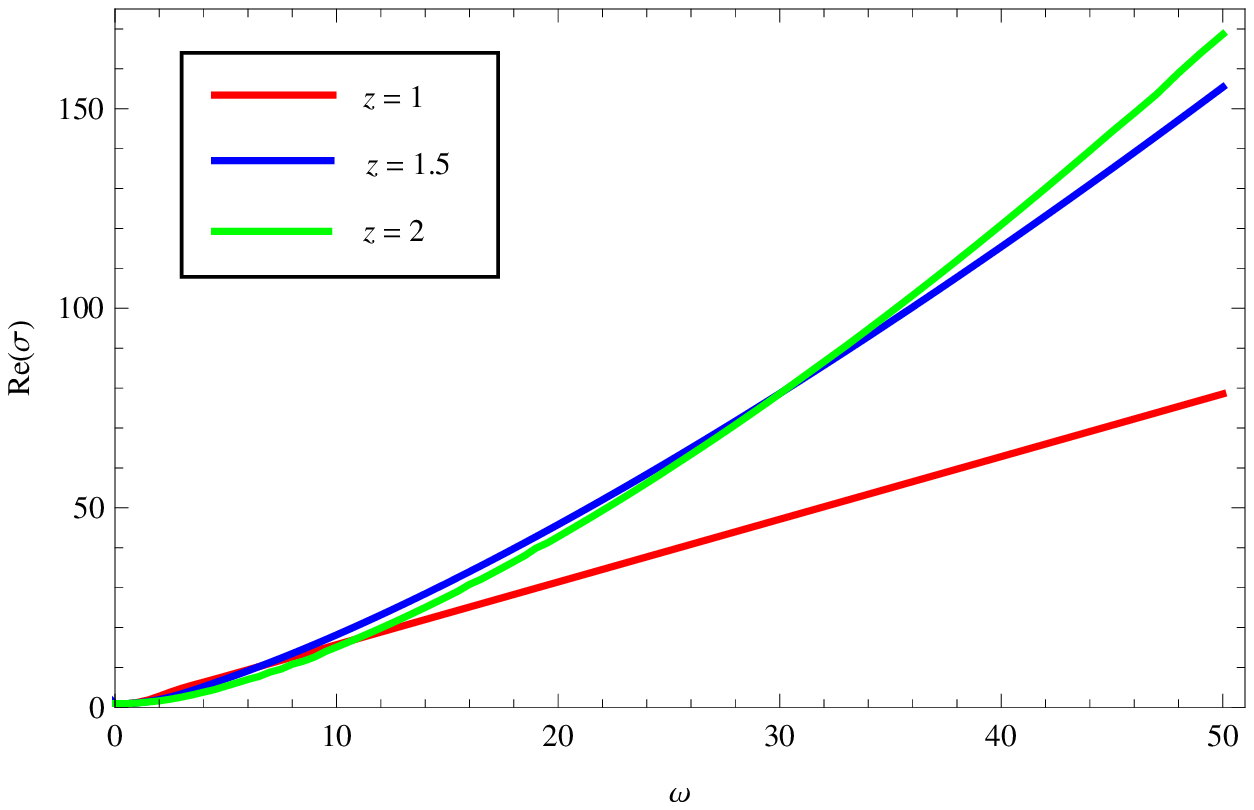}
   \includegraphics[scale=0.6]{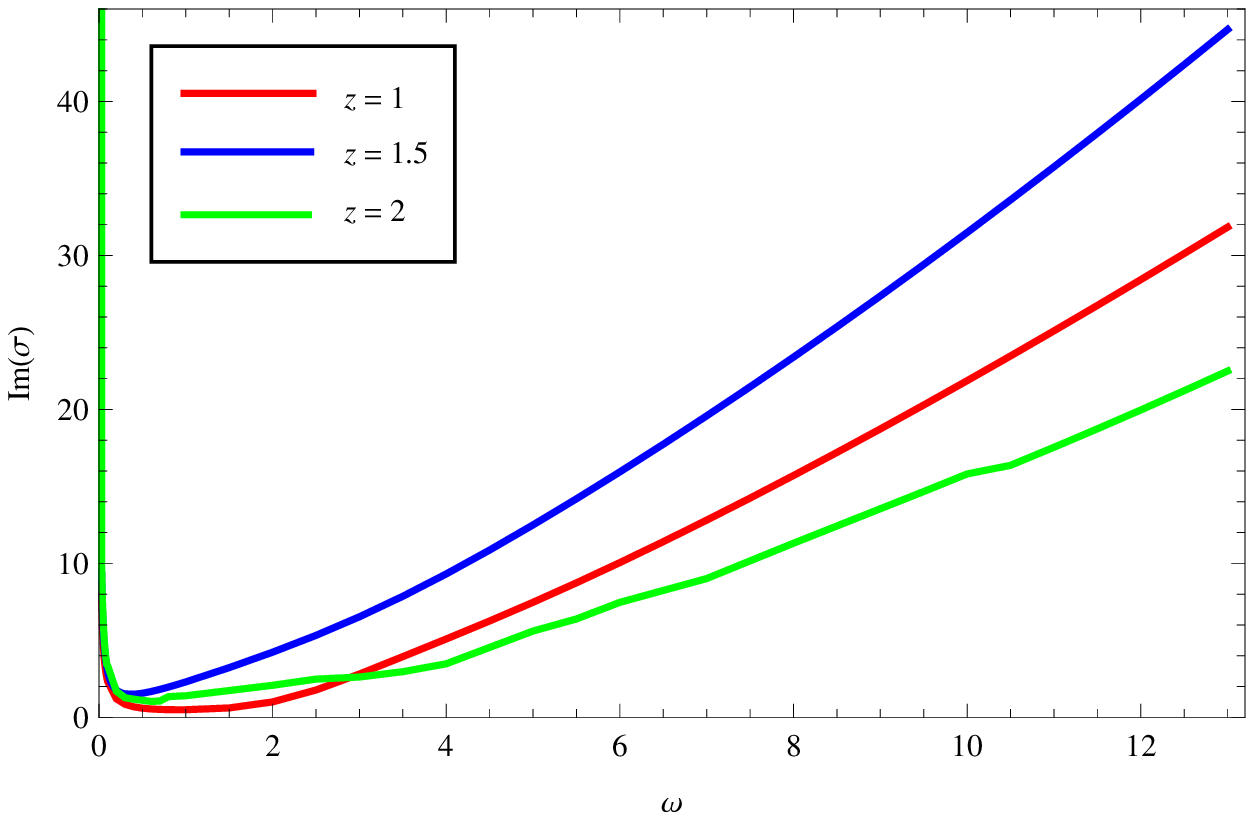}
  \includegraphics[scale=0.6]{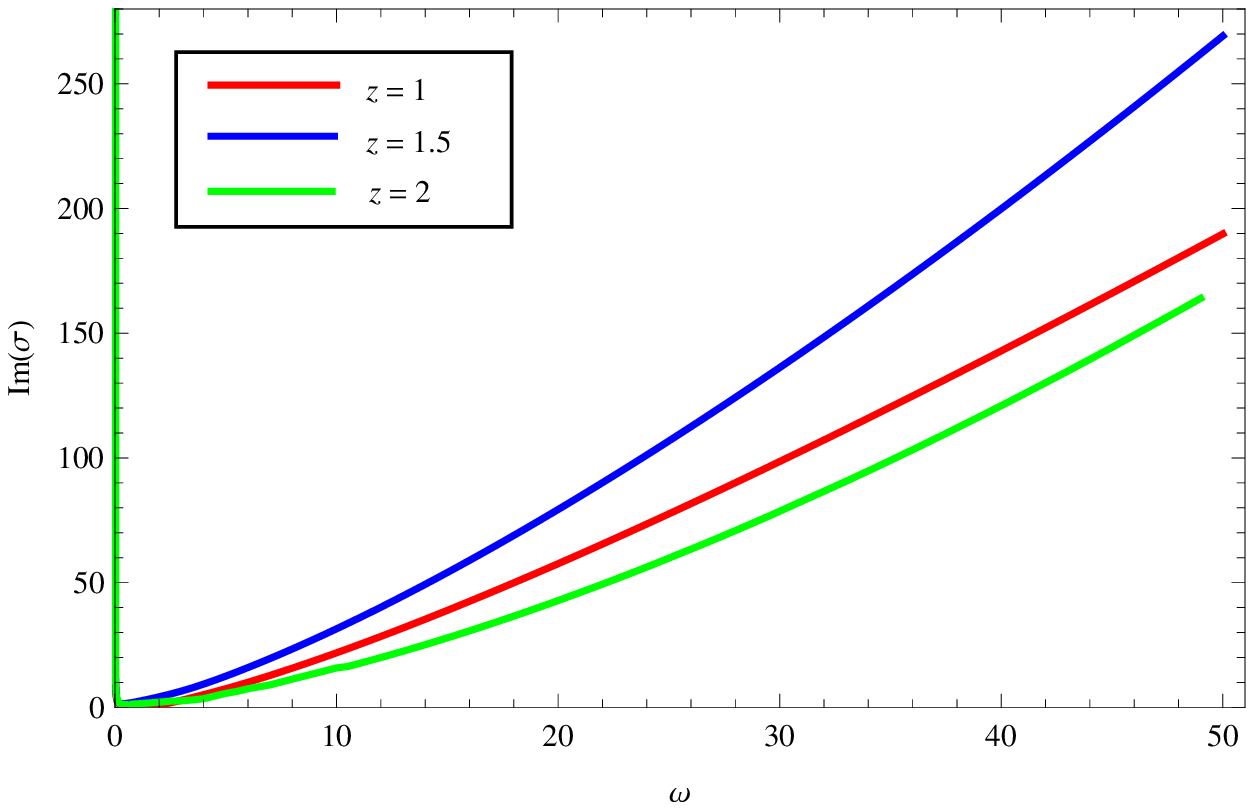}
  \caption{\label{d4}The real and imaginary parts of the conductivity for $d=4$ with respect to various $z$. The left parts are of the low frequency regime while the right parts are of the high frequency regime.}
\end{figure}

From Table \ref{tracoe}, we can find that both of the thermoelectric conductivity $\al$ and the thermal conductivity depend on the electric conductivity $\si$ and the frequency $\om$. Therefore, in Fig.\ref{d3} and Fig.\ref{d4} we only show the numerical results for the electric conductivity $\si$ since the rest transport coefficients can be easily obtained from $\sigma$. In the numerical calculations, we have scaled
$l=1, r_h=1$, and $\rho_2=\mu=1$.

{Actually, in the numerical calculations, we have set the integration starting point very close to the horizon but not exactly equal to $r_h$, because the coefficients of the eq.\eqref{eoma2x} will diverge at $r=r_h$, and we have adopted the usual incoming wave boundary conditions near the horizon.}  From Fig.\ref{d3} and Fig.\ref{d4}, we can find that at $\omega=0$, the real parts of the conductivity is finite; however, the imaginary parts of the conductivity will diverge at $\omega=0$, thus from the Kramers-Kronig relations we can readily deduce that the real parts actually will develop a delta function at $\omega=0$. This delta function is due to the translational invariance of the system. These are known in previous literatures \cite{Hartnoll:2009sz}.

For large frequencies, the expansions for $a_{2x}$ can be found in Table \ref{expansion} in which the coefficients $C_2$ and $C_1$ are functions of $\omega$. Therefore, from Table \ref{tracoe} as well as Table \ref{expansion}, we can get the approximate behavior of the conductivity depending on the frequency $\omega$ as,
\be\label{largeom}
 \sigma_{d=3}(\omega)\sim\left\{
  \begin{array}{cl}
    \omega^0, & {z=1;} \\
    \omega^{2/3}, & {z=3/2;} \\
    \omega(a+\log(\om)), & {z=2.}
  \end{array}
\right.
\quad
\sigma_{d=4}(\omega)\sim\left\{
  \begin{array}{cl}
    \omega(b+\log(\om)), & {z=1;} \\
    \omega^{4/3}, & {z=3/2;} \\
    \omega^{3/2}, & {z=2.}
  \end{array}
\right.
\ee
where, $a$ and $b$ are some constants. This large frequency behavior of the conductivities can be seen from the right parts of Fig. \ref{d3} and Fig. \ref{d4}.

In Fig.\ref{d3},  the real part of the conductivity will tend to a constant when $\omega$ becomes large for $z=1$, which is similar to the previous papers \cite{Hartnoll:2008kx,Hartnoll:2009sz}. But the differences are in the case of $z=3/2$ and $z=2$, in which the Re($\sigma$) will depend on $\omega$ according to eq.\eqref{largeom}. This is an interesting and new phenomenon from the viewpoint of the gauge/gravity duality, which is not observed before in the previous literatures as far as we know. For example, in \cite{Hartnoll:2009sz} the author argued that the electric conductivity in the case of $d=3$ will tend to a constant because of the dimensional analysis. However, here we can explicitly see that in our model for $d=3$ and $z>1$, $\sigma$ will be proportional to $\omega^{s(z)}$ in the large frequency limit, where $s$ is a function of $z$. This peculiar frequency dependent AC electric conductivity may be related to some new materials in the realistic world. Fortunately, in \cite{Dyre:2000zz} the author has studied the AC conductivity for various disordered solids in $(d=2+1)$ and $(d=3+1)$ dimensions both experimentally and theoretically. We found that the electric conductivity for $d=3$ and $d=4$ in our Fig.\ref{d3} and Fig.\ref{d4} have similar behaviors to the experiments or the computer simulations in the large frequency limit in the paper \cite{Dyre:2000zz}. In that paper, the author has proposed a kind of `symmetric hopping model' to illustrate the large frequency behavior of the electric conductivities.  Therefore, we expect that the Lifshitz black brane model in the present paper may be related to this kind of `symmetric hopping model' from certain aspects. We will further report this kind of relation in another work \cite{swz}.

In Fig.\ref{d4} for $z=1$, the large frequency behavior of the conductivity is like $\omega(b+\log(\om))$, which resembles the expansions in the Appendix in \cite{Horowitz:2008bn}. The arguments for the conductivity for $z=3/2$ and $z=2$ are the same as those for $d=3$ in Fig. \ref{d3}.

\section{The Shear viscosity}
\label{sect:shear}
 As we know that any interacting field theory at finite temperature in the limit of  long time  and long wavelength  can be effectively described by hydrodynamics. In this   section, we will compute the shear viscosity of the dual field theory in the low frequency limit. To do so, we need to turn on the transverse tensor mode fluctuation (which is the scalar channel) of the metric $\dl g_{\mu\nu}=h_{xy}$.

\subsection{The case of $N=1$}
Let us begin with the $N=1$ case first, see eq.\eqref{N=1}. Taking the mode expansion of the
fluctuation $\delta g_{xy}(t,r)=h_{xy}(r)e^{-i\omega t+ik\zeta}$ (where $\zeta=x_{d-1}$, see the Appendix),
we obtain the linearized Einstein equation of the $xy$ component  as,
\begin{eqnarray}\label{scalar}
\varphi''+\left(\frac{z+d}{r}+\frac{\Xi'}{\Xi}\right)\varphi'
+\left(\frac{l^{2+2z}\omega^2}{r^{2+2z}\Xi^2}-\frac{k^2l^4}{r^4\Xi}\right)\varphi=0.
\end{eqnarray}
which is the equation of motion of a  minimally coupled
massless scalar  field propagating in the unperturbed spacetime background,
where we have defined $\varphi=h^{x}_{y}$.

To solve eq.(\ref{scalar}), it is convenient to introduce the new
coordinate $u^{2}=\frac{r^{z+d-1}_h}{r^{z+d-1}}$, then the boundary is located at $u=0$,
while $u=1$ is the horizon. After taking the long wavelength limit $k^2\rightarrow 0$, the fluctuation equation becomes
\begin{eqnarray}\label{eomN1}
\varphi''+\left(\frac{\widetilde\Xi'}{\widetilde\Xi}-\frac{1}{u}\right)\varphi'+\frac{4l^{2(z+1)}\omega^{2}u^{\frac{2(z-d+1)}{d+z-1}}}{r^{2z}_h(z+d-1)^{2}\widetilde\Xi^{2}}\varphi=0.
\end{eqnarray}
where $\widetilde\Xi(u)=1-u^{2}$ and ``$'$" is the derivative with respect to $u$. At the horizon, since we are going to calculate the retarded Green's function of the dual field theory, we need to impose the incoming wave boundary
condition. Thus, we set $\varphi=(1-u)^{\alpha}\Psi(u)$, then $\alpha$ can be determined through the near horizon expansion of eq.(\ref{eomN1}), which gives $\alpha=-\frac{i\omega}{r^{z}_h(z+d-1)}$. To obtain the solution of $\Psi(u)$ in the full spacetime region, we can expand it in terms of $\omega$ as
\be
\Psi(u)=\Psi_{0}(u)+\omega
\Psi_{1}(u)+\mathcal{O}(\omega^2),\ee
and then solve the above equation order by order. Furthermore, requiring $\Psi_{0}$ to be regular at the horizon and normalizing it to be one
at the boundary,  as well as $\Psi_{1}$ vanishes at the horizon,  we find that
\be
\Psi_{0}=1 \quad {\rm and}\quad
\Psi_{1}=-\frac{i}{(z+d-1)r^{z}_h}\ln(\frac{1+u}{2}),\ee
then we have
\be
\varphi=(1-u)^{-\frac{i\omega}{r^{z}_h(z+d-1)}}\left(1-\frac{i\omega}{(z+d-1)r^{z}_h}\ln(\frac{1+u}{2})\right).
\ee

To compute the shear viscosity of the boundary field theory, we need to compute the flux factor
$\mathcal{F}=K\sqrt{-g}g^{uu}\varphi^\ast(u)\partial_{u}\varphi(u)$,
where $K$ is a normalization constant related to the effective coupling constant of the bulk transverse graviton. Keeping to the order of $O(\omega)$, it
is straightforward to compute the flux factor and the retarded 2-point Green's
function  as
\be G_{R}=-2\mathcal{F}|_{u=0}=-\frac{i\omega r^{d-1}_h}{16\pi
G_{d+1}l^{d-1}},\ee
so the shear viscosity  can be obtained by the Kubo formula as
\begin{equation}
\eta=-\lim_{\omega\rightarrow0}\frac{{\rm Im}
G_{R}(\omega,\vec{k}=0)}{\omega}=\frac{r^{d-1}_h}{16\pi
G_{d+1}l^{d-1}},
\end{equation}
then we have
\be \frac{\eta}{s}=\frac{1}{4\pi} \ee
which satisfies the  KSS bound in the Einstein gravity.

\subsection{The case of $N\geq 2$}

Now we consider $N\geq 2$ cases, see eq.\eqref{bgd}. As we have shown in the Appendix, the equation of motion for $\varphi=h^{x}_{y}$ is also that
of a minimally coupled massless scalar field, which is of the same
form  as eq.(\ref{scalar})
\begin{equation}\label{scalar2}
\varphi''(r)+\left(\frac{f'}{f}+\frac{d+z}{r}\right)\varphi'(r)+
\left(\frac{l^{2z+2}\omega^2}{r^{2z+2}f^2}-\frac{l^4 k^2}{r^4
f}\right)\varphi(r)=0.
\end{equation}
Note that since $f(r)$ has multiple zero roots and cannot be determined in general, to solve eq.(\ref{scalar2}), it is more convenient to apply the matching method in which the exact form of $f(r)$ is not involved.

In the near horizon region, i.e. $r-r_h\ll r_h$, $f(r)\simeq
f'(r_h)(r-r_h)$, then eq.(\ref{scalar2}) can be simplified as
\be\label{scalar3} \varphi''(r)+\frac{1}{r-r_h}\varphi'(u)+
\left(\frac{c_1\omega^{2}}{(r-r_h)^2}-\frac{c_2k^2}{r-r_h}\right)\varphi(r)=0,\ee
in which
\be c_1=\left(\frac{1}{4\pi T}\right)^2\quad {\rm and}\quad
c_2=\frac{1}{4\pi T}\left(\frac{r_h}{l}\right)^{z-3}.\ee
Let's further defining $\bar{r}=r/r_h$ and taking the long
wavelength limit $k^2\rightarrow0$, eq.(\ref{scalar3}) becomes
\be\label{scalar4}
\varphi''(\bar{r})+\frac{1}{\bar{r}-1}\varphi'(\bar{r})+
\frac{c_1\omega^{2}}{(\bar{r}-1)^2}\varphi(\bar{r})=0,\ee
which gives
\be\label{nh} \varphi(\bar{r})=\bar{c_3}(\bar{r}-1)^{\frac{i
\mathfrak{w}}{2}}+\bar{c_4}(\bar{r}-1)^{\frac{-i \mathfrak{w}}{2}},\ee
in the $r$ coordinate the solution is
\be\label{nh1} \varphi(r)=c_3(r-r_h)^{\frac{i
\mathfrak{w}}{2}}+c_4(r-r_h)^{\frac{-i \mathfrak{w}}{2}},\ee
where $\mathfrak{w}=\frac{\omega}{2\pi T}$. The first part of \eqref{nh} or \eqref{nh1} is the outgoing mode
while the second part is the ingoing mode. To calculate the retarded
Green's function, we need to adopt the ingoing mode, which require
$\bar{c_3}=c_3=0$ in eq.(\ref{nh}) and eq.\eqref{nh1}. In the low frequency limit,
eq.(\ref{nh1}) can be expanded as
\be\label{nh2} \varphi(r)=c_4\left(1-\frac{i\omega}{4\pi
T}\ln(r-r_h)+\mathcal{O}(\omega^2)\right),\ee

In the near region, $r_h\omega<r\omega\ll 1$, then in the
$k^2\rightarrow0$ limit, eq.(\ref{scalar2}) reduces to
\be \label{scalar7}
\varphi''(r)+\left(\frac{f'}{f}+\frac{d+z}{r}\right)\varphi'(r)=0,\ee
which can be solved as
\be\label{near} \varphi(r)=\int\frac{c_5}{f r^{d+z}}dr+c_6.\ee
Note that in the near horizon limit $r\rightarrow r_h$,
eq.(\ref{near}) can be simplified as
\be\label{near1} \varphi(r)\approx\int
\frac{c_5}{f'(r_h)(r-r_h)r_h^{d+z}}dr +
c_6=\frac{c_5}{f'(r_h)r_h^{d+z}}\ln(r-r_h)+c_6.\ee
While in the large radius limit, $f(r)\rightarrow 1$, then
eq.(\ref{near}) becomes
\be\label{far1}
\varphi(r)\approx\int\frac{c_5}{r^{d+z}}dr+c_6=-\frac{c_5}{(d+z-1)}\frac{1}{r^{d+z-1}}+c_6,\ee

In the outer region $r_h\ll l\ll r$, $f'(r)\rightarrow 0$,
$f(r)\rightarrow 1$ and again we taking $k^2\rightarrow 0$, then
eq.(\ref{scalar2}) becomes
\begin{equation}\label{scalar8}
\varphi''(r)+\frac{d+z}{r}\varphi'(r)+
\frac{l^{2z+2}}{r^{2z+2}}\omega^2\varphi(r)=0,
\end{equation}
in the $u={1}/{r}$ coordinate, eq.(\ref{scalar8}) can be changed to
\begin{equation}\label{scalar82}
\varphi''(u)-\frac{d+z-2}{u}\varphi'(u)+
l^{2z+2}u^{2z-2}\omega^2\varphi(u)=0,
\end{equation}
ant its solution is
\be\label{out} \varphi(u)=u^{\frac{\Delta_+}{2}}\left(c_7
J_{-\frac{\Delta_+}{2z}}\left(\frac{l^{1+z}\omega u^z}{z}\right)+c_8
J_{\frac{\Delta_+}{2z}}\left(\frac{l^{1+z}\omega
u^z}{z}\right)\right),\ee
where
\be
c_7=\bar{c_7}(2z)^{-\frac{\Delta_+}{2z}}(l^{1+z}\omega)^{\frac{\Delta_+}{2z}}\Gamma\left(\frac{1-d+z}{2z}\right)\quad
{\rm and}\quad
c_8=\bar{c_8}(2z)^{-\frac{\Delta_+}{2z}}(l^{1+z}\omega)^{\frac{\Delta_+}{2z}}\Gamma\left(\frac{-1+d+3z}{2z}\right)\nno
\ee
in which, $\bar c_7$ and $\bar c_8$ are certain constants while $\Delta_+=d+z-1$ is the conformal dimension of the operator dual
to the massless scalar field in the bulk. Again, in the low
frequency limit, eq.(\ref{out})  can be expanded as
\be\label{out1}\varphi(r)=\bar{c_7}\left(1+\mathcal{O}(\omega^2)\right)
+\bar{c_8}l^{\frac{(1+z)\Delta_+}{z}}\left(\frac{2}{z}\right)^{\frac{\Delta_+}{z}}
\omega^{1+\frac{d-1}{z}}r^{-\Delta_+}\left(1+\mathcal{O}(\omega^2)\right).\ee

The condition for matching the solutions in these three regions is
$r_h< r\ll \omega^{-1}$. Comparing the eq.(\ref{near1}) with
eq.(\ref{nh2}) we get that
\be c_4=c_6 \quad {\rm and}\quad -i\omega
c_4=\frac{c_5}{l^{z+1}r_h^{d-1}}.\ee
While the matching of eq.(\ref{far1}) with eq.(\ref{out1}) gives
\be c_6=\bar{c_7} \quad {\rm and}\quad
-\frac{c_5}{d+z-1}=\bar{c_8}l^{\frac{(1+z)\Delta_+}{z}}
\left(\frac{2}{z}\right)^{\frac{\Delta_+}{z}}\omega^{1+\frac{d-1}{z}}.\ee
Namely, the coefficients in  these three regions are related by the following
relations
\be \bar{c_7}=c_6=c_4 \quad {\rm and}\quad
\bar{c_8}l^{\frac{(1+z)\Delta_+}{z}}
\left(\frac{2}{z}\right)^{\frac{\Delta_+}{z}}\omega^{1+\frac{d-1}{z}}
=c_4\frac{l^{z+1}r_h^{d-1}}{d+z-1}i\omega.\ee
Furthermore, the normalization condition requires that $\varphi(r)$
is normalized to be $1$, namely, $c_4=1$. Consequently, the asymptotic
solution at low frequency limit becomes
\be\label{out2}\varphi(r)=\left(r-r_h\right)^{\frac{-i
{\mathfrak{w}}}{2}}\left(1+\frac{l^{z+1}r_h^{d-1}r^{-\Delta_+}}{(d+z-1)}i\omega
+\mathcal{O}(\omega^2)\right) .\ee
After eliminating the divergent terms, the dominant part of the
radial flux of the scalar field at the boundary is
\be
\mathcal{F}&=&K\sqrt{-g}g^{rr}\varphi^*(r)\partial_{r}\varphi(r)|_{r\rightarrow \infty}\nno\\
&=&-i K \frac{r_h^{d-1}}{l^{d-1}}\omega +\mathcal{O}(\omega^2)\nno\\
&=&-\frac{i}{32\pi
G_{d+1}}\frac{r_h^{d-1}}{l^{d-1}}\omega+\mathcal{O}(\omega^2),\ee
where $K=1/(32\pi G_{d+1})$ is the effective coupling constant of
the scalar field $\varphi(r)$, then the retarded Green's function is
\be G_R(k)=-2\mathcal{F}(k,r)|_{r\rightarrow \infty},\ee
and the shear viscosity is calculated from the Kubo formula
\be \eta &=&-\lim_{\omega\rightarrow 0}\frac{{\rm Im}
G_R}{\omega}\nno\\
&=&\frac{1}{16\pi G_{d+1}}\frac{r_h^{d-1}}{l^{d-1}},\ee
Therefore, the ratio of the shear viscosity to the entropy is
\be\label{eta/s} \frac{\eta}{s}=\frac{1}{4\pi},\ee
which gives the same value as that of the Lifshitz black brane with
only one $U(1)$ gauge field. The result
indicates that the additional background $U(1)$ gauge fields do not alter the KSS bound of the boundary fluid, though they do contribute to the shear viscosity and the entropy density, respectively.

\section{Conclusions and discussions}
\label{sect:conclusion}
In this paper, we  studied the model of strongly coupled non-relativistic quantum field theory with multiple $U(1)$ gauge fields near the Lifshitz fix points, in the frame work of the non-relativistic gauge/gravity duality. By considering the linearized perturbations of bulk gravitational and gauge fields, we solved the equation of motions for gauge fields with back reactions (shear channel) and the bulk transverse graviton (scalar channel). For the $N=2$ case, we derived the renormalized second order effective action and systematically calculated the electric, thermal and thermoelectric conductivities of the dual non-relativistic quantum field theories with respect to various $d$ and $z$. Specifically, we found the novel frequency dependent power law behavior of the AC electric conductivity in the large frequency limit when $d=3$ and $z>1$. From the knowledge of the condensed matter physics, we expect that our model provides a holographic description of the `symmetric hopping model' in some sense. The argument goes to the case of $d=4$ as well, we will report the further relationship between the Lifshitz black brane and the hopping conductivities in another paper elsewhere. In addition, when taking the limit of long wavelength and low frequency in the generic $N$ cases, we also showed that the ratio of shear viscosity to entropy density of the dual boundary fluids still satisfies the KSS bound derived in the Einstein gravity.

\section*{Acknowledgement}
We would like to thank Jeppe C. Dyre, Sean Hartnoll for kind response and Da-Wei Pang and Yi Yang for valuable discussions. J.R.S. was supported by the National Science Foundation of China under Grant No. 11147190 and 11205058. S.Y.W. was supported by the National Science Council (NSC 101-2112-M- 009-005 and NSC 101-2811-M-009-015) and National Center for Theoretical Science, Taiwan. H.Q.Z. was supported by a Marie Curie International Reintegration Grant PIRG07-GA-2010-268172.

\begin{appendix}
\section{Linearized Perturbations of the Gravitational Theory}
\subsection{Einstein-Maxwell-dilaton theory}
The Einstein-Maxwell-dilaton theory with multiple $U(1)$
gauge fields that we are considering has the action
\be\label{EMd} I=\frac{1}{16\pi
G_{d+1}}\int{d^{d+1}x}\sqrt{-g}\left(R-2\Lambda-\frac{1}{2}\partial^{\mu}\phi\partial_{\mu}\phi-\frac{1}{4}\sum^{N}_{a=1}e^{\lambda_a\phi}F^{2}_{a}\right),
\ee
its Einstein equation is
\be
R_{\mu\nu}-\frac{2\Lambda}{d-1}g_{\mu\nu}=\frac{1}{2}\partial_{\mu}\phi\partial_{\nu}\phi
+\frac{1}{2}\sum_{a=1}^N e^{\lambda_a
\phi}\left(F_{a\lambda\mu}F_{a\nu}
^{\lambda}-\frac{1}{2(d-1)}F_a^{2}g_{\mu\nu}\right).\ee

Let us make the metric ansatz to be a $d+1$ dimensional black brane solution as
\be \label{BB}ds^2=H_1(r)\left(-f(r)dt^2+dx^i
dx_i\right)+H_2(r)dr^2,\ee
where its outter horizon is located at $f(r_h)=0$.

Consider the small metric fluctuation caused by some external
perturbation
\be g^{(0)}_{\mu\nu}\rightarrow
g_{\mu\nu}=g^{(0)}_{\mu\nu}+\delta g_{\mu\nu},\ee
the Christoffel symbol is
\be
\Gamma^{\lambda}_{\mu\nu}&=&\Gamma^{(0)\lambda}_{\mu\nu}+\delta\Gamma^{\lambda}_{\mu\nu}\nno\\
&=&\Gamma^{(0)\lambda}_{\mu\nu}+\frac{g^{\lambda\alpha}}{2}\left(\nabla_\mu\delta g_{\al\nu}+\nabla_\nu \delta g_{\mu\al}-\nabla_\al \delta g_{\mu\nu}\right),\ee
when taking the linear order perturbation of the metric, i.e. $\delta g_{\mu\nu}=h_{\mu\nu}$, the Christoffel symbol can be expanded up to second order of $h$ as
\be
\Gamma^{\lambda}_{\mu\nu}
&=&\Gamma^{(0)\lambda}_{\mu\nu}+\Gamma^{(1)\lambda}_{\mu\nu}+\Gamma^{(2)\lambda}_{\mu\nu},\ee
where
\be
\Gamma^{(1)\lambda}_{\mu\nu}&=&\frac{g^{(0)\lambda\alpha}}{2}\left(\nabla_\mu h_{\al\nu}+\nabla_\nu h_{\mu\al}-\nabla_\al h_{\mu\nu}\right),\nno\\
\Gamma^{(2)\lambda}_{\mu\nu}&=& -\frac{h^{\lambda\alpha}}{2}\left(\nabla_\mu h_{\al\nu}+\nabla_\nu h_{\mu\al}-\nabla_\al h_{\mu\nu}\right).\ee
Note that under the first order variation, the Ricci tensor varies as
\be
R_{\mu\nu}&=&R^{(0)}_{\mu\nu}+\delta R^{(0)}_{\mu\nu}\nno\\
&=&R^{(0)}_{\mu\nu}+R^{(1)}_{\mu\nu}+R^{(2)}_{\mu\nu}
,\ee
where
\be R^{(1)}_{\mu\nu}&=&\Gamma^{(1)\alpha}_{\mu\nu;\alpha}-\Gamma^{(1)\alpha}_{\mu\alpha;\nu}\nno\\
&=&\frac 1 2 (\nabla^\alpha \nabla_\mu
h_{\alpha\nu}+\nabla^\alpha \nabla_\nu h_{\alpha\mu})-\frac
1 2 \Box h_{\mu\nu}-\frac 1 2 \nabla_\nu \nabla_\mu h,\ee
and
\be  R^{(2)}_{\mu\nu}&=&\Gamma^{(2)\alpha}_{\mu\nu;\alpha}-\Gamma^{(2)\alpha}_{\mu\alpha;\nu}\nno\\&=&-\frac{h^{\al\beta}}{2}\left(\nabla_\al\nabla_\mu h_{\beta\nu}+\nabla_\al\nabla_\nu h_{\beta\mu}-\nabla_\al\nabla_\beta h_{\mu\nu}\right)+\frac{h^{\al\beta}}{2}\nabla_\nu\nabla_\mu h_{\al\beta}\nno\\
&&-\frac{\nabla_\al h^{\al\beta}}{2}\left(\nabla_\mu h_{\beta\nu}+\nabla_\nu h_{\beta\mu}-\nabla_\beta h_{\mu\nu}\right)+\frac{\nabla_\nu h^{\al\beta}}{2}\nabla_\mu h_{\al\beta},\ee
the first order and second order Ricci scalars are
\be
R^{(1)}&=&g^{(0)\mu\nu}R^{(1)}_{\mu\nu}
-h^{\mu\nu}R^{(0)}_{\mu\nu}\nno\\
&=& \nabla^\alpha \nabla^\beta h_{\alpha\beta}-\Box h -
\frac{2\Lambda}{d-1}h,\ee
and
\be R^{(2)}&=&g^{(0)\mu\nu}R^{(2)}_{\mu\nu}
-h^{\mu\nu}R^{(1)}_{\mu\nu}\nno\\
&=&-h^{\mu\nu}\left(\nabla_\mu\nabla^\la h_{\nu\la}+\nabla^\la\nabla_\mu h_{\nu\la}\right)+h^{\mu\nu}\nabla_\mu\nabla_\nu h+h^{\mu\nu}\Box h_{\mu\nu}\nno\\
&&-\nabla_\al h^{\al\la}\nabla^\beta h_{\beta\la}+\frac{\nabla_\al h^{\al\la}}{2}\nabla_\la h+\frac{\nabla_\la h^{\mu\nu}}{2}\nabla^\la h_{\mu\nu},\ee
respectively.

Then the linearized Einstein equation is
\be\label{lEinsteineq2}
R^{(1)}_{\mu\nu}-\frac{2\Lambda}{d-1}h_{\mu\nu}=\frac{1}{2}\sum_{a=1}^N
e^{\lambda_a \phi}\left(-F_{a\alpha\mu}F_{a\beta\nu}
h^{\alpha\beta}-\frac{1}{2(d-1)}(F_a^{2}h_{\mu\nu}-2F_{a\alpha}
^{\gamma}F_{a\gamma\beta}h^{\alpha\beta}g^{(0)}_{\mu\nu})\right)\ee
When there is only transverse gravitational fluctuation
$h_{xy}=h_{xy}(r)e^{-i\omega t+ik\zeta}$, where $\zeta=x_{d-1}$ is the $d-1$-th spatial coordinate. Using the $R^{(0)x}_{x}$ component of the zeroth order eom, i.e.
\be R^{(0)x}_{x}-\frac{2\Lambda}{d-1}=\frac{1}{2}\sum_{a=1}^N
e^{\lambda_a \phi}\left(-
\frac{1}{2(d-1)}F_{a\lambda\alpha}F_{a\gamma\beta}g^{(0)\lambda\gamma}g^{(0)\alpha\beta}\right),\nno\ee
then eq.(\ref{lEinsteineq2}) becomes
\be\label{lEinsteineq3} -\frac{H_1}{2}\Box\varphi
+\left(\frac{H_1^{'2}}{2H_1H_2}-\frac 1 2 \Box
H_1\right)\varphi=\frac{2\Lambda}{d-1}H_1\varphi-\frac{1}{2}\sum_{a=1}^N
e^{\lambda_a
\phi}\frac{1}{2(d-1)}F_a^{2}H_1\varphi=R^{(0)}_{xx}\varphi\ee
which gives the eom of minimally coupled massless scalar field $\varphi=h^x_y$
\be -\frac{H_1}{2}\Box\varphi
=-\frac{H_1}{2}\frac{1}{\sqrt{-g^{(0)}}}\partial_\mu\left(\sqrt{-g^{(0)}}g^{(0)\mu\nu}\partial_\nu\varphi\right)=0.\ee
%


\subsection{Gauge field perturbation with back reaction}
To compute the conductivities of the dual field theory, we
need to turn on the gauge field perturbation along the spatial
direction, this gauge field perturbation will in turn induce the
$h_{ti}$ off-diagonal part of the background metric perturbation
since $h_{ti}$ and $a_{ai}$ are the vector mode fluctuations. Without loss of
generality, we choose $\delta
A_{a\mu}=\delta^{\lambda}_{x}a_{a\lambda}(r)e^{-i\omega t+ik\zeta}$,
which induces the corresponding metric perturbation is $\delta
g_{\mu\nu}=h_{tx}(r)e^{-i\omega t+ik\zeta}$. Then the linearized
Einstein and Maxwell equations are obtained by making the combined
diffeomorphism and gauge variations to the original equations,
namely
\be
\delta_{\epsilon+\chi}\left(R_{\mu\nu}-\frac{2\Lambda}{d-1}g_{\mu\nu}\right)=\delta_{\epsilon+\chi}\left(\frac{1}{2}\partial_{\mu}\phi\partial_{\nu}\phi
+\frac{1}{2}\sum_{a=1}^N e^{\lambda_a
\phi}\left(F_{a\lambda\mu}F_{a\nu}
^{\lambda}-\frac{1}{2(d-1)}F_a^{2}g_{\mu\nu}\right)\right),\ee
where $\delta_\epsilon$ means the diffeomorphism transformation
while $\delta_\chi$ indicates the gauge field transformation that
obeying the following relations
\be \delta_\epsilon g_{\mu\nu}=\mathcal{L}_\epsilon g_{\mu\nu}\quad
{\rm and}\quad \delta_\chi A_{a\mu}=a_{a\mu}.\ee
In the linear order perturbation, the nonvanishing components of the
first order Ricci tensor are $R_{xt}=R_{tx}$ and $R_{xr}=R_{rx}$. Then
the linearized Einstein equation are
\be\label{lEinsteineq4}
R^{(1)}_{xt}-\frac{2\Lambda}{d-1}h_{xt}&=&\frac{1}{2}\sum_{a=1}^N e^{\lambda_a
\phi}\left(g^{(0)rr}\partial_r a_{ax}\partial_r
A_{at}-\frac{1}{2(d-1)}F^{(0)2}_a h_{xt}\right),\ee
together with the $xx$ component of the zeroth order Einstein equation, eq.(\ref{lEinsteineq4}) becomes
\be &&\frac{1}{4fH_1 H_2^2}\left(H_2 f'(h_{tx}' H_1-h_{tx} H_1')+fH_1(h_{tx}' H_2'-2h_{tx}'' H_2)+f h_{tx}(-H_1' H_2'+2H_1'' H_2)\right)\nno\\
&&=\frac{1}{2H_2}\sum_{a=1}^N
e^{\lambda_a \phi}a_{ax}' A_{at}',\ee
and
\be R^{(1)}_{rt}&=&-\frac{1}{2}\sum_{a=1}^N e^{\lambda_a \phi}g^{(0)tt}\partial_t
a_{a x}A_{a t}' = \frac{i\omega}{2}\sum_{a=1}^N e^{\lambda_a
\phi}g^{(0)tt}a_{a x}A_{a t}',\ee
which gives
\be\label{lEM} h'_{tx}-\frac{H_1'}{H_1}h_{tx}+\sum_{a=1}^N
e^{\lambda_a \phi}a_{a x}A'_{a t}=0,\ee
where ``$'$'' indicates $\partial_r$.

In addition, the linearized Maxwell equation is obtained by
\be
\label{Maxwell}&&\delta_{\epsilon+\chi}\partial_{\mu}\left(\sqrt{-g}e^{\lambda_a
\phi}g^{\mu\alpha}g^{\nu\beta}F_{a\alpha\beta}\right)\nno\\
&=&\partial_{\mu}\left(\frac{\sqrt{-g^{(0)}}}{2}e^{\lambda_a
\phi}g^{(0)\rho\sigma}h_{\rho\sigma}g^{(0)\mu\alpha}g^{(0)\nu\beta}F^{(0)}_{a\alpha\beta}\right)\nno\\
&&-\partial_{\mu}\left(\sqrt{-g^{(0)}}e^{\lambda_a
\phi}(h^{\mu\alpha}g^{(0)\nu\beta}+g^{(0)\mu\alpha}h^{\nu\beta})F^{(0)}_{a\alpha\beta}\right)\nno\\
&&+\partial_{\mu}\left(\sqrt{-g^{(0)}}e^{\lambda_a
\phi}g^{(0)\mu\alpha}g^{(0)\nu\beta}(\partial_\alpha
a_{a\beta}-\partial_\beta a_{a\alpha})\right)=0,\ee
its nonvanishing components are
\be \partial_r\left(\sqrt{-g}e^{\lambda_a
\phi}g^{(0)rr}g^{(0)xx}g^{(0)tt}(-h_{tx})\partial_r
A_{at}\right)&+&\partial_t\left(\sqrt{-g}e^{\lambda_a
\phi}g^{(0)xx}g^{(0)tt}\partial_t
a_{ax}\right)\nno\\&+&\partial_r\left(\sqrt{-g}e^{\lambda_a
\phi}g^{(0)rr}g^{(0)xx}\partial_r a_{ax}\right)=0.\nno\ee
In the black brane background eq.(\ref{BB}), the above equations
become
\be\label{lM}
a''_{ax}+\left(\frac{(d-2)H'_1}{2H_1}-\frac{H'_2}{2H_2}+\frac{f'}{2f}+\lambda_a
\phi' \right)a'_{ax}+\frac{\omega^2H_2
}{fH_1}a_{ax}=\left(\frac{H'_1h_{tx}}{fH_1^2}-\frac{h'_{tx}}{fH_1}\right)A'_{at}.\ee

When taking the ansatz $H_1=b^2$, $H_2=1/\xi$, $f(r)=\xi e^{-\chi}/b^2$ and
$\ga_a(\phi)=e^{\lambda_a \phi}$ in eq.(\ref{metricansatz}), eq.(\ref{lEM}) and eq.(\ref{lM})
change into
\be h'_{tx}-\frac{2b'}{b}h_{tx}+\sum_{a=1}^N \ga_a(\phi)A'_{a
t}a_{a x}=0,\ee
and
\be
a''_{ax}+\left(\frac{(d-3)b'}{b}+\frac{\xi'}{\xi}-\frac{\chi'}{2}+\frac{\phi'}{\ga_a(\phi)}
\frac{d\ga_a(\phi)}{d\phi}\right)a'_{ax}
+\frac{\omega^2}{\xi^2}e^{\chi}a_{ax}=\left(\frac{2b'h_{tx}}{\xi b}-\frac{h'_{tx}}{\xi}\right)A'_{at}e^{\chi}.\ee
%

In the linear order perturbation of the metric and the gauge fields, the bulk action can also be expanded into second order as
\be I&=&I^{(0)}+I^{(1)}+I^{(2)},\ee
where the zeroth order action is
\be\label{I0} I^{(0)}&=&\frac{1}{16\pi
G_{d+1}}\int{d^{d+1}x}\sqrt{-g^{(0)}}\left(R^{(0)}-2\Lambda-\frac{1}{2}g^{(0)\mu\nu}\partial_{\mu}\phi\partial_{\nu}\phi-\frac{1}{4}\sum^{N}_{a=1}e^{\lambda_a\phi}F^{(0)2}_{a}\right)\nno\\
&=&\frac{1}{16\pi
G_{d+1}}\int{d^{d+1}x}\sqrt{-g^{(0)}}\left(\frac{4\Lambda}{d-1}-\frac{1}{2(d-1)}\sum^{N}_{a=1}e^{\lambda_a\phi}F^{(0)2}_{a}\right),
\ee
and the first order action is
\be\label{I1} I^{(1)}&=&\frac{1}{16\pi
G_{d+1}}\int{d^{d+1}x}\sqrt{-g^{(0)}}\left(\Na^\mu\Na^\nu h_{\mu\nu}-\Box h-\frac{1}{2}\sum^{N}_{a=1}e^{\lambda_a\phi}F^{(0)}_{a\mu\nu}F^{(1)\mu\nu}_{a}\right)\nno\\
&=&\frac{1}{16\pi
G_{d+1}}\int_\Si{d^{d}x}\sqrt{-g^{(0)}}n_\mu\left(\Na^\nu h^\mu_{\nu}-\Na^\mu h-\sum^{N}_{a=1}a_{a\nu}\left(e^{\lambda_a\phi}F^{(0)\mu\nu}_{a}\right)\right),\ee
which are purely surface terms when the bulk eoms are satisfied (on-shell condition), where $n_\mu$ is the unit normal vector of the hypersurface $\Si$.

While the second order action is
\be\label{I2} I^{(2)}&=&\frac{1}{16\pi
G_{d+1}}\int{d^{d+1}x}\sqrt{-g^{(0)}}\{-h^{\mu\nu}\Na^\la\Na_\mu h_{\nu\la}+\frac 1 2 h^{\mu\nu}\Na_\mu\Na_\nu h+\frac 1 2 h^{\mu\nu}\Box h_{\mu\nu} \nno\\
&&-\frac{1}{4}\sum^{N}_{a=1}e^{\lambda_a\phi}\left(F^{(1)}_{a\mu\nu}F^{(1)\mu\nu}_{a}
-4F^{(1)}_{a\al\la}F^{(0)\la}_{a\beta}h^{\al\beta}+F^{(0)}_{a\mu\al}F^{(0)}_{a\nu\beta}h^{\mu\nu}h^{\al\beta}\right)\nno\\
&&+\left(\frac 1 2 h^{\mu\nu}h_{\mu\nu}+\frac 1 4 h^2\right)\mathcal{L}^{(0)}
+\frac{h}{2}\mathcal{L}^{(1)}\}\nno\\
&&+\frac{1}{16\pi
G_{d+1}}\int_\Si{d^{d}x}\sqrt{-g^{(0)}}n_\la\left(-h^{\la\nu}\Na^\mu h_{\mu\nu}+\frac 1 2 h^{\la\mu}\Na_\mu h+\frac 1 2 h^{\mu\nu}\Na^\la h_{\mu\nu}\right),\ee
where $\mathcal{L}^{(0)}$ and $\mathcal{L}^{(1)}$ are respectively the first and second order Lagrangian densities in $I^{(0)}$ and $I^{(1)}$.

\end{appendix}


\end{document}